\documentclass[usenatbib,letterpaper]{mn2e}
\usepackage{times,amssymb,epsfig,aas_macros}
\usepackage{longtable}

\usepackage{color}
\definecolor{dr}{rgb}{0.6,0,0}
\definecolor{db}{rgb}{0,0,0.6}

\usepackage[ hyperfootnotes={false},
dvips,
pdftitle={Observing Kepler stars with WISE},
pdfauthor={Grant Kennedy},
pdfstartview={FitH},
bookmarksopen={true} ]{hyperref}
\hypersetup{colorlinks={true}, urlcolor={db}, citecolor={dr}}

\begin{document}

\title[Observing \emph{Kepler} stars with WISE]{Confusion limited surveys: using WISE to
  quantify the rarity of warm dust around \emph{Kepler} stars}

\author[G. M. Kennedy, M. C. Wyatt]
{G. M. Kennedy\thanks{Email:
    \href{mailto:gkennedy@ast.cam.ac.uk}{gkennedy@ast.cam.ac.uk}} \& M. C. Wyatt \\
  Institute of Astronomy, University of Cambridge, Madingley Road, Cambridge CB3 0HA,
  UK\\
}

\maketitle

\begin{abstract}
  We describe a search for infra-red excess emission from dusty circumstellar material
  around 180,000 stars observed by the \emph{Kepler} and WISE missions. This study is
  motivated by i) the potential to find bright warm disks around planet host stars, ii) a
  need to characterise the distribution of rare warm disks, and iii) the possible
  identification of candidates for discovering transiting dust concentrations. We find
  about 8,000 stars that have excess emission, mostly at 12$\mu$m. The positions of these
  stars correlate with the 100$\mu$m background level so most of the flux measurements
  associated with these excesses are spurious. We identify 271 stars with plausible
  excesses by making a 5MJy/sr cut in the IRAS 100$\mu$m emission. The number counts of
  these excesses, at both 12 and 22$\mu$m, have the same distribution as extra-Galactic
  number counts. Thus, although some excesses may be circumstellar, most can be explained
  as chance alignments with background galaxies. The one exception is a 22$\mu$m excess
  associated with a relatively nearby A-type star that we were able to confirm because
  the disk occurrence rate is independent of stellar distance. This detection implies a
  disk occurrence rate consistent with that found for nearby A-stars. Despite our low
  detection rate, these results place valuable upper limits on the distribution of large
  mid-infrared excesses; e.g. fewer than 1:1000 stars have 12$\mu$m excesses ($F_{\rm
    obs}/F_\star$) larger than a factor of five. In contrast to previous studies, we find
  no evidence for disks around 1790 stars with candidate planets (we attribute one
  significant 12$\mu$m excess to a background galaxy), and no evidence that the disk
  distribution around planet hosts is different to the bulk population. Higher resolution
  imaging of stars with excesses is the best way to rule out galaxy confusion and
  identify more reliable disk candidates among \emph{Kepler} stars. A similar survey to
  ours that focusses on nearby stars would be well suited to finding the distribution of
  rare warm disks.
\end{abstract}

\begin{keywords}
  planets and satellites: formation --- stars:individual, HD 69830, BD +20 307, HD
  172555, $\eta$ Corvi, HIP 13642, KIC 7345479
\end{keywords}

\section{Introduction}\label{s:intro}

\emph{Kepler} \citep{2003SPIE.4854..129B} is revolutionising our perspective on
extra-Solar planets
\citep[e.g.][]{2010Sci...330...51H,2011Natur.470...53L,2011ApJ...729...27B,2011Sci...333.1602D,2011arXiv1112.2165H,2012ApJ...745..120B}
and will likely yield many Earth-sized planets in the terrestrial zones of their host
stars. Like the Solar System, these planetary systems will comprise not only planets, but
also smaller objects that for one reason or another did not grow larger. In the Solar
System these make up the Asteroid and Kuiper belts, along with other populations such as
the Oort cloud, and Trojan and irregular satellites. Characterisation of these
populations has been critical to building our understanding of how the Solar System
formed. For example, one of the primary validation methods of the so-called ``Nice''
model for the origin of the outer Solar System's architecture has been the reproduction
of these minor body populations and their properties
\citep[e.g.][]{2005Natur.435..462M,2008Icar..196..258L,2007AJ....133.1962N}.

The exquisite detail with which the Solar System minor body populations are characterised
is made clear when they are contrasted with their extra-Solar analogues, collectively
known as ``debris disks.'' First discovered around Vega \citep{1984ApJ...278L..23A}, they
are almost always detected by unresolved infra-red (IR) emission, visible as an excess
above the stellar photosphere. Detection of an excess at multiple wavelengths yields the
dust temperature, and thus the approximate radial distance from the star (to within a
factor of a few). The radial location can be refined further when spectral features are
present \citep[e.g.][]{2007ApJ...658..584L}. However, because the temperature of a dust
or ice grain depends on size, the true radial location (and any radial, azimuthal, or
vertical structure) can generally only be found by resolved imaging
\citep[e.g.][]{1984Sci...226.1421S,2005Natur.435.1067K} or interferometry
\citep[e.g.][]{2006A&A...452..237A,2009A&A...503..265S}.

It is therefore difficult to draw links between the regions of planetary systems occupied
by planets and small bodies, and if and how they interact. The best examples of
extrasolar systems where known dust and planets are likely to interact are $\beta$
Pictoris, Fomalhaut, and HR 8799
\citep{1995AAS...187.3205B,1997MNRAS.292..896M,2005Natur.435.1067K,2008Sci...322.1345K,2008Sci...322.1348M,2009ApJ...705..314S,2010ApJ...717.1123M}. In
these cases, the spatial dust distribution is fairly well known because the disk is
resolved, but the orbits of the planets, which were only discovered recently with direct
imaging, are not. These are rare cases however, and typically the search for links
between the major and minor body components of extra-Solar planetary systems means asking
whether the presence of one makes the presence of the other more or less likely. So far
no statistically significant correlation between the presence of planets and debris has
been found \citep{2009ApJ...700L..73K,2009ApJ...705.1226B,2011AJ....141...11D}. However,
there is new tentative evidence that nearby stars with low-mass planetary systems are
more likely to harbour debris than those with no planet detections (Wyatt et al, in
press), an exciting possibility that has only been achievable recently with better
sensitivity to such planetary systems around nearby stars.


One of the key limiting factors in the search for links between debris and planets is the
small number of stars known to host both. Two recent \emph{Spitzer} surveys observed
about 150 planet host stars, of which about 10\% were found to have disks
\citep{2009ApJ...705.1226B,2011AJ....141...11D}. The small number of disk detections is
therefore the product of the number of nearby stars known to host planets that could be
observed with \emph{Spitzer}, and the $\sim$10\% disk detection rate (for both planet and
non-planet host stars). One way to sidestep this problem is therefore to look for disks
around a much larger sample of planet host stars; the \emph{Kepler} planet host
candidates \citep{2011ApJ...736...19B,2012arXiv1202.5852B}.

The method we use to look for disks in this study is to find infra-red (IR) excess
emission above that expected from the stellar photosphere. An IR excess is usually
interpreted as being thermal emission from an Asteroid or Kuiper-belt analogue, which is
heated by the star it orbits. We use photometry from the Wide-field Infrared Survey
Explorer (WISE) mission's \citep{2010AJ....140.1868W} all-sky catalogue, which is most
sensitive to dust in the terrestrial region of Sun-like stars. Three properties of warm
dust at these relatively close radial distances provide motivation.

First, this warm dust, if discovered, is located in the vicinity of the planets being
discovered with \emph{Kepler}. Currently, only one system, HD 69830, is known to host
both a planetary system in the terrestrial region and warm dust
\citep{2005ApJ...626.1061B,2006Natur.441..305L}. The origin of this dust is unclear, but
given the proximity to the planetary system is plausibly related
\citep{2007ApJ...658..584L,2011ApJ...743...85B}. Through discovery of similar systems the
links between planets and warm dust can be better understood.

For planets discovered by \emph{Kepler}, the knowledge that a transiting planetary system
is almost exactly edge-on provides the second motivational aspect. If planets pass in
front of the host star, so will coplanar minor body populations. Indeed, the discovery of
systems where multiple planets transit their stars
\citep[e.g.][]{2010Sci...330...51H,2011Natur.470...53L,2011ApJS..197....8L} provides
striking evidence that the Solar System's near-coplanar configuration is probably
typical. While transits of individual small bodes will be impossible to detect, it may be
possible to detect concentrated populations that arise from a recent collision
\citep{2005AJ....130..269K} or perturbations by planets \citep{2011AJ....142..123S}. The
dust must reside on a fairly close orbit---within a few AU---to allow multiple transits
within the mission lifetime. Thus, the WISE sensitivity to terrestrial dust, and likely
difficulties in discerning dust transits from other instrumental and real effects, mean
that the odds of finding dust transits might be maximised by the prior identification of
dusty systems.

Finally, but most importantly, detections of terrestrial dust are rare
\citep[e.g.][]{1991ApJ...368..264A,2006ApJ...638.1070H,2006ApJ...636.1098B,2006ApJ...639.1166B}. Because
only a few such systems are known, their occurrence rate is poorly constrained. More
discoveries are therefore needed to add to our understanding of the processes that create
it. The collision rate in a debris disk is proportional to the orbital period, so warm
terrestrial debris disks decay to undetectable levels rapidly, hence their
rarity. Indeed, the few that are known are usually thought to be the result of recent
collisions, and thus transient phenomena \citep[e.g. HD 69830, HD 172555, BD +20 307,
$\eta$ Corvi, HD 165014, HD 169666, HD
15407A][]{2005ApJ...626.1061B,2005Natur.436..363S,2007ApJ...658..569W,2009ApJ...701.2019L,2009ApJ...700L..25M,2010ApJ...714L.152F,2011arXiv1110.4172L,2012ApJ...749L..29F}.
Possible scenarios include objects thrown into the inner regions of a planetary system
from an outer reservoir
\citep{1999ApJ...510L.131G,2007ApJ...658..569W,2009MNRAS.399..385B,2011A&A...530A..62R,2012MNRAS.420.2990B},
or the remnant dust from a single catastrophic collision
\citep{2005Natur.436..363S,2005ApJ...626.1061B,2011ApJ...726...72W}.

Clearly, there are reasons that discovery of debris in the terrestrial regions of known
planetary systems is important. However, because WISE is sensitive to the rarest and
brightest disks around \emph{Kepler} stars, the third point above is of key
importance. As stars become more distant, they and their debris disks become fainter, and
the number of background galaxies at these fainter flux levels increases. Thus, the bulk
of the stars in the \emph{Kepler} field, which lie at distances of hundreds to thousands
of parsecs may not be well suited to debris disk discovery. Practically, the importance
of contamination depends on the galaxy contamination frequency relative to the disk
frequency (i.e. only if disks are too rare will they be overwhelmed by
contamination). \emph{Therefore, because the occurrence rate of the rare disks that WISE
  is sensitive to is unknown, whether \emph{Kepler} stars are a good sample for disk
  detection with WISE is also unknown.}

Characterising the occurrence rate of rare bright disks is therefore the main goal of
this study, because this very distribution sets what can be discovered. While the sample
of \emph{Kepler} stars is not specifically needed for this goal, there is the possibility
that the disk occurrence rate is higher for stars that host low-mass planets. Such a
trend could make this particular sub-sample robust to confusion, even if the general
population is not.

An additional potential issue specific to the \emph{Kepler} field is the importance of
the Galactic background. High background regions are sometimes avoided by debris disk
observations because they make flux measurement difficult, and can even mask the presence
of otherwise detectable emission. Unfortunately this issue cannot be avoided for the
present study, as the \emph{Kepler} field is necessarily located near the Galactic plane
to maximise the stellar density on the sky.

In what follows, we describe our search for warm excesses around $\sim$180,000 stars
observed by \emph{Kepler} using the WISE catalogue. We first outline the data used in
this study in \S \ref{s:cat} and in \S \ref{s:xs} describe our SED fitting method for
finding excesses and the various issues encountered. We discuss the interpretation of
these excesses in \S \ref{s:interp}, and place our findings in the context of disks
around nearby stars in \S \ref{s:nearby}. We discuss the disk-planet relation, rarity of
warm bright excesses, and some future prospects in \S \ref{s:disc} and conclude in \S
\ref{s:sum}. Readers only interested in the outcome of this search may wish to skip the
details described in \S\S \ref{s:cat}-\ref{s:xs}.

\section{Catalogues}\label{s:cat}

\begin{figure*}
  \begin{center}
    \hspace{-0.5cm} \includegraphics[width=0.5\textwidth]{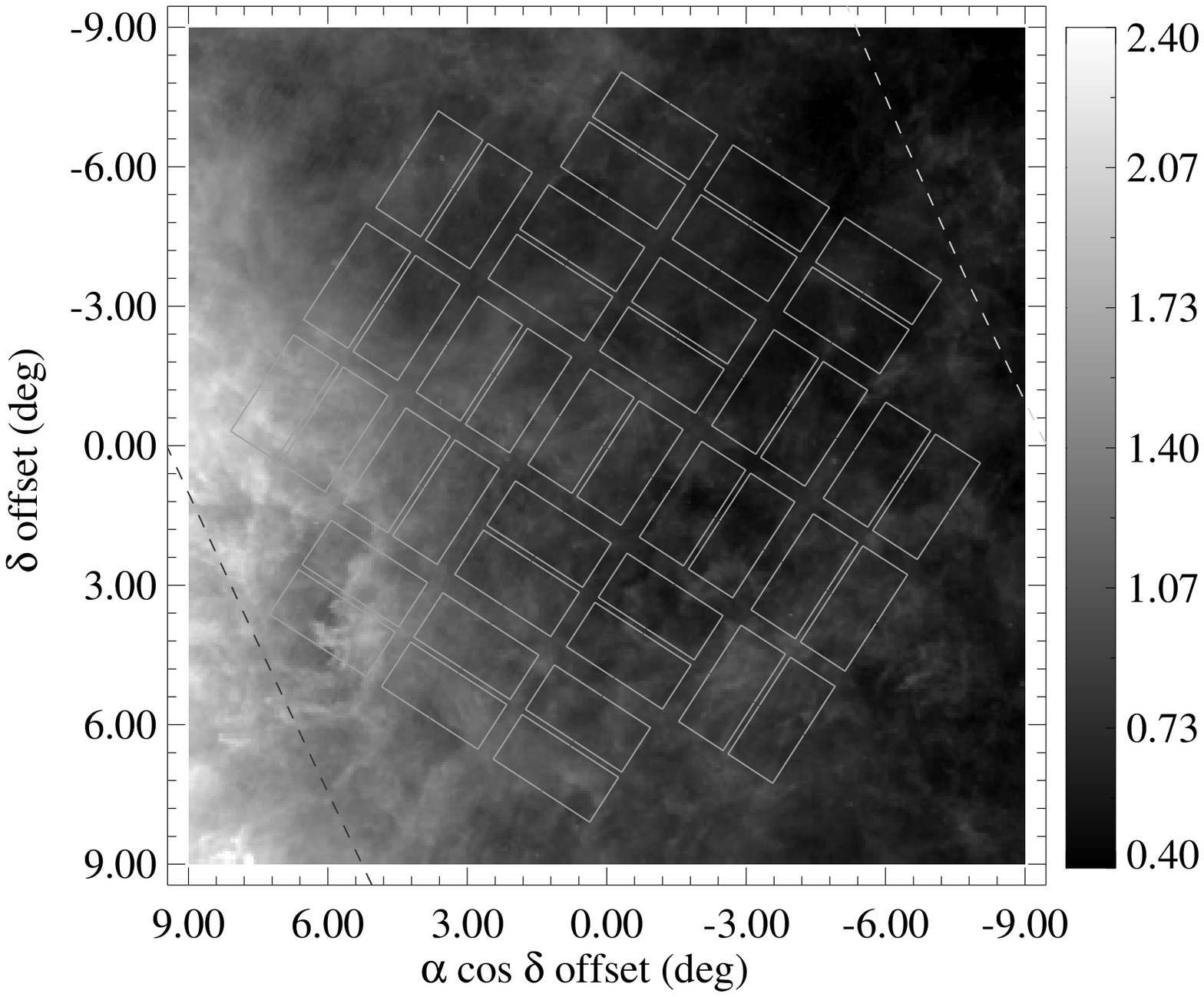}
    \hspace{0.25cm} \includegraphics[width=0.5\textwidth]{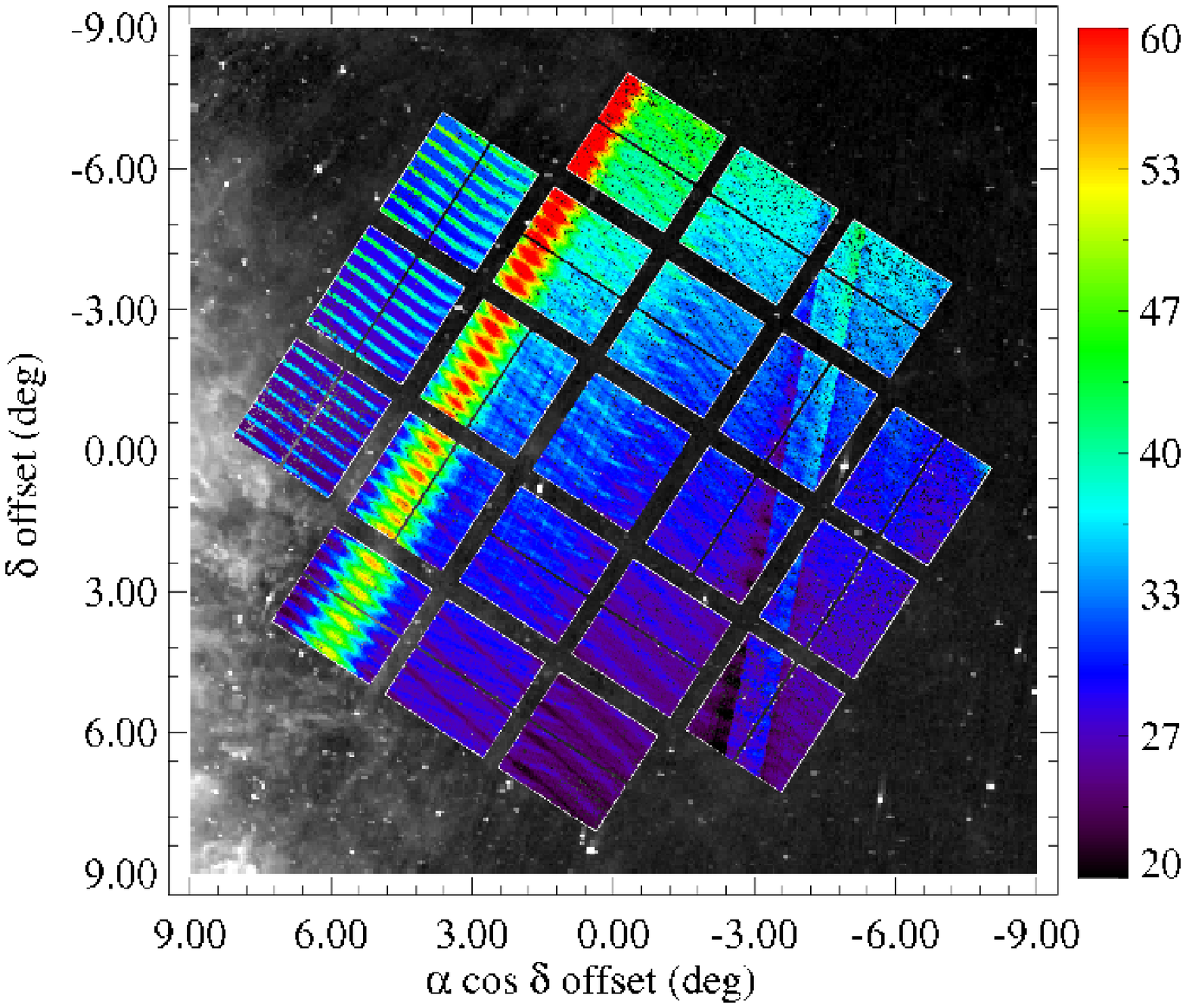}
    \caption{Kepler field of view (rectangles) with IRAS 100$\mu$m (left panel, scale is log
      MJy/sr) and 25$\mu$m (right panel) IRIS maps as background
      \citep{2005ApJS..157..302M}. North is up and East is left. The Galactic plane is
      towards the South-East, and in the left panel galactic latitudes of 5 and
      22$^\circ$ are shown as dashed lines. The WISE W1 mean pixel coverage in frames
      (called ``w1cov'' in the catalogue, coloured dots and scale) is shown in the right
      panel. Nearly all objects are observed by 20-60 frames, those with less or more
      have the colour at the respective end of the colour scale.}\label{fig:w1cov}
  \end{center}
\end{figure*}

The \emph{Kepler} mission is observing $\sim$200,000 stars near the Galactic plane to
look for planets by the transit method \citep{2003SPIE.4854..129B}. The \emph{Kepler}
field of view (FOV) covers about 100 square degrees and is rotated by 90$^\circ$ every
three months. Not all stars are observed in all quarters, but Figure \ref{fig:w1cov}
shows that the focal plane has four-fold rotational symmetry (aside from the central
part), so most will be visible for the mission lifetime. Stars observed by \emph{Kepler}
are brighter than about 16th magnitude and selected to maximise the chance of transit
detection and follow up \citep{2010ApJ...713L.109B}. The stars are drawn from the Kepler
Input Catalogue (KIC), which contains optical photometry, cross-matched 2MASS IDs, and
stellar parameters for millions of objects within the FOV \citep{2011AJ....142..112B}.

The entire \emph{Kepler} FOV is covered by the all-sky WISE mission
\citep{2010AJ....140.1868W}, as shown by the coloured dots in the right panel of Figure
\ref{fig:w1cov}. The scanning strategy used by WISE means that the sky coverage varies,
and is highly redundant at the ecliptic poles \citep[see][]{2011ApJ...735..112J}. WISE
photometry comprises four bands with isophotal wavelengths of 3.4, 4.6, 12, and 22$\mu$m
(called W1-4). The sensitivity is fairly well suited to stars observed by \emph{Kepler},
with 5$\sigma$ sensitivities of 17.1, 15.7, 11.4, and 8 magnitudes for 8 frames in W1-4
(corresponding to 44, 93, 800, and 5500$\mu$Jy respectively).

\subsection{Cross matching}\label{ss:xc}

There were 189,998 unique KIC objects observed in quarters 1-6 of the mission, which we
refer to as Kepler OBserved objects,\footnote{Retrieved from
  \href{http://archive.stsci.edu/pub/kepler/catalogs/}{http://archive.stsci.edu/pub/kepler/catalogs/}}
or ``KOBs''.  These KOBs are matched with three photometric catalogues: Tycho 2
\citep{2000A&A...355L..27H}, the 2MASS Point Source Catalogue
\citep{2006AJ....131.1163S}, and the WISE all-sky catalogue
\citep{2010AJ....140.1868W}. Matching with 2MASS is straightforward because designations
are already given in the KIC. The relevant 189,765 rows were retrieved using
Vizier.\footnote{\href{http://vizier.u-strasbg.fr/viz-bin/VizieR}{http://vizier.u-strasbg.fr/viz-bin/VizieR}}
Tycho 2 objects were matched using a 1'' search radius and retaining only the closest
object, again using the Vizier service. This match returned 13,430 objects. The KIC
itself contains photometry in SDSS-like bands ($ugriz$ each with 1092, 189383, 189829,
186908, 177293 KOB measurements respectively) and a DDO51-like narrowband filter centered
on 510nm (with 176170 measurements). Finally, WISE objects are matched using the IPAC
Gator
service\footnote{\href{http://irsa.ipac.caltech.edu/Missions/wise.html}{http://irsa.ipac.caltech.edu/Missions/wise.html}}
with a radius of 1'', which returns 181,004 matches. It is these 181,004 objects that are
the focus of this study.

\section{Finding excesses}\label{s:xs}

The method used to identify debris disk candidates is fitting stellar atmospheric model
spectral energy distributions (SEDs) to the available photometry (known as ``SED
fitting''). Optical and near-IR bands are used to fit the stellar atmosphere and make
predictions of the photospheric flux at longer wavelengths, which are compared to the
WISE observations. An IR excess indicates the possible presence of thermal emission from
circumstellar dust, but can also arise for other reasons; overestimated flux due to a
high Galactic background, chance alignment with a background galaxy that has a cooler
spectrum than a star, and poor photospheric prediction are three examples. Excesses are
usually quantified in two ways: the first is the flux ratio in a band $B$
\begin{equation}\label{eq:rx}
  R_{\rm B} = F_{\rm B} / F_{\star {\rm B}},
\end{equation}
where $F_{\rm B}$ is the photometric measurement and $F_{\star {\rm B}}$ is the
photospheric prediction. The flux from the disk is therefore
\begin{equation}\label{eq:fdisk}
  F_{\rm disk,B} = \left( R_{\rm B} - 1 \right) F_{\star {\rm B}}.
\end{equation}
The second is the excess significance,\footnote{This quantity is usually called usually
  called $\chi_{\rm B}$, but we instead dub it $X_{\rm B}$ to avoid confusion with the
  goodness of fit indicator $\chi^2$. In this study we usually use the term ``excess
  significance'' instead of the symbol.}
\begin{equation}\label{eq:chix}
  X_{\rm B} = \frac{ F_{\rm B} - F_{\star {\rm B}} }
  { \sqrt{ \sigma_{\rm B}^2 + \sigma_{\star {\rm B}}^2 } } ,
\end{equation}
where each $\sigma$ is the photometric or stellar photospheric uncertainty. Typically a
star is said to have excess emission when $X_{\rm B} > 3$, though other (usually
higher) values appropriate to the sample in question may be used.

\begin{figure}
  \begin{center}
    \hspace{-0.5cm} \includegraphics[width=0.5\textwidth]{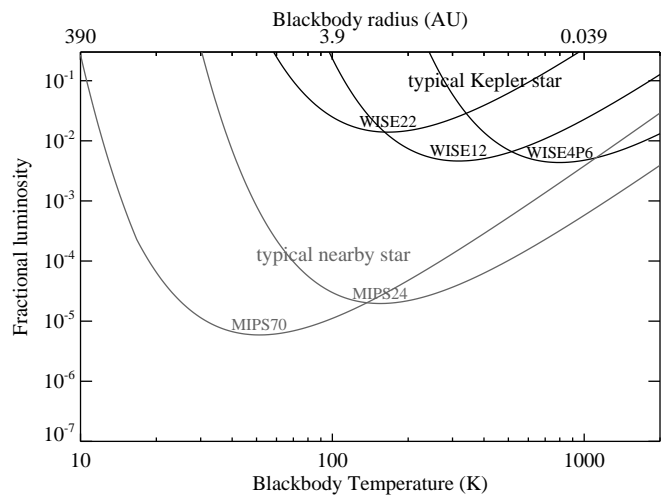}
    \caption{WISE W2-4 (4.6-22$\mu$m) 3$\sigma$ sensitivity to blackbody emission around
      a typical star in the \emph{Kepler} field, a 5460K star with $K_s=13.5$mag (black
      lines). Also shown is the sensitivity with MIPS at 24 and 70$\mu$m to 55 Cancri, a
      typical nearby star at 12pc with $K_s \approx 4.2$ (grey lines).}\label{fig:sens}
  \end{center}
\end{figure}

Figure \ref{fig:sens} illustrates the sensitivity of WISE to blackbody emission around a
``typical'' KOB; a 5460K star with $K_s=13.5$mag. Only disks that have a combination of
fractional luminosity ($f = L_{\rm disk}/L_\star$) and temperature that lies above the
line for a specific band can be detected in that band. Disks detected at a single
wavelength lie somewhere along a single curve similar to (but above) those shown, and may
be constrained by non-detections at other wavelengths \citep[e.g. Figure 9
of][]{2006ApJ...636.1098B}. Temperatures can only be derived for disks that lie above
multiple lines (i.e. are detected at multiple wavelengths). Compared to the MIPS
observations of the nearby (12pc) star 55 Cancri \citep{2008ApJ...674.1086T}, the WISE
sensitivity is much reduced. While MIPS 24$\mu$m observations are generally ``calibration
limited'' by the precision of the stellar photospheric predictions (i.e. $\sigma_{\star
  {\rm B}} \gg \sigma_{\rm B}$), WISE can at best detect a flux ratio of about 50 at
22$\mu$m for the example shown here \citep[so is ``sensitivity limited'' and
$\sigma_{\star {\rm B}} \ll \sigma_{\rm B}$, see][for further discussion of sensitivity
vs. calibration limited surveys]{2008ARA&A..46..339W}. The WISE 22$\mu$m observation is
actually slightly deeper than the MIPS 24$\mu$m one (both are sensitive to about 1mJy),
so the reason for the difference in disk sensitivity is simply the brightness of the
star. The radial scale shows that WISE is well suited to finding excesses with large
fractional luminosity that lie within the terrestrial planet zone of Sun-like stars. To
detect dust at larger distances would require either much greater sensitivity and/or
longer wavelength data.

Because the sensitivity depends strongly on the brightness of the star, it also varies
widely for KOBs. For the nearest and brightest stars the sensitivity is about two orders
of magnitude better than shown for the KOB in Figure \ref{fig:sens}. However, the longest
WISE wavelength is 22$\mu$m, so even for the brightest stars the sensitivity to disks
cooler than 100K drops significantly. Therefore, regardless of brightness, the wavelength
range of WISE and the brightness of most KOBs means that any detections will be due to
very luminous warm excesses. It is less likely that the Wein side of cooler emission will
be detected because much higher fractional luminosities are required (i.e. the curves in
Figure \ref{fig:sens} rise very steeply to cooler temperatures and larger radii).

\subsection{SED fitting}\label{ss:sed}

Our SED fitting method uses filter bandpasses to compute synthetic photometry and colour
corrections for the stellar models, which are fit to observed photometry by a combination
of brute force grids and least-squares minimisation. We use Phoenix AMES-Cond models from
the Gaia grid \citep{2005ESASP.576..565B}, which cover a wide range of stellar
parameters. However, these models only have $T_{\rm eff}<10,000$K, so stars pegged at
this temperature are re-fit with Kurucz models \citep{2003IAUS..210P.A20C}. Previous
efforts for the \emph{Herschel} \citep{2010A&A...518L...1P} Disc Emission via a Bias-free
Reconnaissance in the Infrared/Submillimetre (DEBRIS) Key Programme
\citep[e.g.][]{2010A&A...518L.135M} have found that there is little or no difference
between these models in terms of photospheric predictions for A-stars.

In addition to being near the Galactic plane, most stars observed by \emph{Kepler} are
hundreds to a few thousands of parsecs distant, so are reddened by interstellar dust. The
100 and 25$\mu$m IRIS maps\footnote{Retrieved from
  \href{http://skyview.gsfc.nasa.gov/}{http://skyview.gsfc.nasa.gov/}.} in Figure
\ref{fig:w1cov} show that cool emission from dust generally increases towards the
Galactic plane, but also varies on scales much smaller than the \emph{Kepler} FOV. We
correct for this effect using the \citet{1985ApJ...288..618R} reddening law.

There are five possible stellar parameters to include in the SED fitting: the effective
temperature ($T_{\rm eff}$), surface gravity ($\log g$), metallicity ([M/H]), reddening
($A_V$), and the solid angle of the star ($\Omega_\star$). The stellar radius $R_\star$
can subsequently be estimated from $A_V$ by adopting some model for Galactic reddening
\citep[e.g.][]{2011AJ....142..112B} or by assuming the star has a specific luminosity
class. In early SED fitting runs where all parameters were left free, models were
commonly driven to implausible regions of parameter space in order to minimise the
$\chi^2$. Similar issues lead \citet{2011AJ....142..112B} to use a Bayesian approach,
with priors based on observed stellar populations. Rather than duplicate and/or verify
their method, we use some of their KIC stellar parameters as described below because our
goal is to obtain the best photospheric prediction in WISE bands, not to derive stellar
parameters.

\citet{2011ApJ...738L..28V} show that the KIC gravities systematically differ from those
derived by astroseismology by about 0.23dex, though unfortunately the discrepancy is
strongest for those with $\log g > 4$ (i.e. dwarfs).  In order to reduce the number of
fitted parameters, we therefore fix $\log g$ in our fitting to the KIC value minus
0.23dex.  Where no $\log g$ is tabulated, we set it to 4.5, appropriate for the
Solar-type stars that make up the bulk of KOBs. To further reduce the number of
parameters, we also fix [M/H] to the KIC value.

We use least squares minimisation to find an adequately fitting model, starting with the
parameters tabulated in the KIC. When no parameters are given \citep[most likely because
a good fit could not be found for the KIC,][]{2011AJ....142..112B}, we do a grid search
over $T_{\rm eff}$ and $\Omega_\star$ at $A_V=0$ to find an initial fit, and then iterate
to the best fit with these as free parameters (we apply a cut in fit quality below so if
the fits remain poor the stars are excluded).

We use photometric bands up to and including WISE W1 to fit the stellar
atmosphere. Including W1 in the stellar fit is reasonable because main-sequence stars
rarely show excesses shortward of about 10$\mu$m. In fact, bands such as IRAS 12$\mu$m
and AKARI 9$\mu$m can usually be used to fit the photosphere, and it is only in rare
cases, mostly for A-type stars, that these bands show an excess. In the present case
however, the sample is two to three orders of magnitude larger than a typical debris disk
survey, and our goal is to find rare excesses in these bands.

\subsubsection{Discarding suspect and poor photometry}\label{sss:susp}

Care must be taken when using the photometry from the 2MASS and WISE catalogues. The very
large number of sources means that small issues that are usually ignored result in
hundreds of spurious excesses. Both catalogues have various quantities that can be used
to identify and mitigate these problems. These are either flags that indicate
contamination, saturation, upper limits, etc., or values that quantify some property such
as the goodness of fit achieved in the source extraction. Most issues were uncovered in
early SED fitting runs, where the distribution of some catalogue property (source
extraction quality for example) was very different for excesses than for the bulk
population.

\begin{figure}
  \begin{center}
    \hspace{-0.cm} \includegraphics[width=0.5\textwidth]{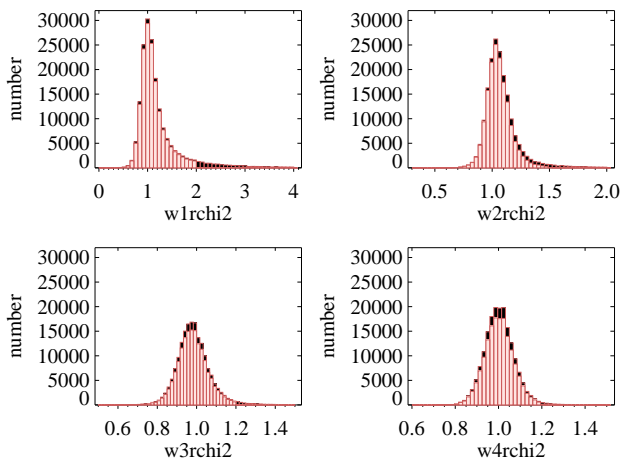}
    \caption{Goodness of fit quality for WISE source extraction in bands W1-4
      (\texttt{w1,2,3,4rchi2}). The non-Gaussian W1 distribution is an indication of
      confusion with nearby sources. The black histograms show all WISE sources, and the
      pink histograms show those remaining after the cut in goodness of fit described in
      the text.}\label{fig:chi2}
  \end{center}
\end{figure}

The WISE catalogue has several indicators that can be used to remove suspect photometry,
which are related to either source extraction or image
artefacts.\footnote{\href{http://wise2.ipac.caltech.edu/docs/release/allsky/expsup/sec2\_2a.html}{http://wise2.ipac.caltech.edu/docs/release/allsky/expsup/sec2\_2a.html}}
We first consider the reduced $\chi^2$ from the source extraction for each WISE bands,
whose distributions are shown in Figure \ref{fig:chi2} (the columns in the catalogue are
called \texttt{wXrchi2} where \texttt{X} is 1, 2, 3, or 4). These measure the quality of
the profile fitting source extraction; a high value is a likely indicator that the source
is not well described by a point source, so could be resolved or confused with another
object. Any deviation from a point source cannot be due to a resolved debris disk; most
\emph{Kepler} stars are hundreds to thousands of parsecs away and WISE is sensitive to
warm excesses that lie at small stellocentric distances (i.e. have very small angular
size)

The most noticeable feature in Figure \ref{fig:chi2} is that W1 shows a non-Gaussian
distribution. Given that the difference in beam FWHM from W1 to W2 is only 6\farcs1
vs. 6\farcs4 and the wavelength difference is small, it seems unlikely that such a large
difference in the $\chi^2$ distributions is astrophysical. However, we found that sources
with poor W1 source extraction were more likely to show excesses (above \texttt{w1rchi2}
of about 2), which we attribute to confusion with nearby sources. Based on this result
and the distributions in Figure \ref{fig:chi2} we avoid poor source extraction by keeping
WISE photometry only when the $\chi^2$ is smaller than 2, 1.5, 1.2, and 1.2 in W1-4
respectively. A similar, but less stringent cut is also made by only retaining sources
where the extension flag (\texttt{ext\_flg}) is zero, which means that no band has a
$\chi^2>3$ and the source is not within 5'' of a 2MASS Extended Source Catalogue entry.

The WISE catalogue also provides flags (the \texttt{cc\_flags} column) that note
contamination from diffraction, persistence, halo, and ghost artefacts. These flags
indicate the estimated seriousness of the contamination; whether the artefact may be
affecting a real source, or the artefact may be masquerading as a source. We avoid
photometry with any indication of contamination (i.e. both types).

Applying these quality criteria to the WISE data results in 126743, 128610, 78340, and
9790 detections with a signal to noise ratio of greater than three in W1-4
respectively. A total of 144655 sources have a WISE detection in at least one band.

The 2MASS catalogue also has columns for the profile fitting source extraction $\chi^2$
in each band (called
\texttt{J,H,Kpsfchi}).\footnote{\href{http://www.ipac.caltech.edu/2mass/releases/allsky/doc/sec4\_4b.html}{http://www.ipac.caltech.edu/2mass/releases/allsky/doc/sec4\_4b.html}}
Early SED fitting runs found that nearly half of the W1-2 excesses had a 2MASS reduced
$\chi^2>2$, particularly in the $J$ band. Given that only about 5\% of all 2MASS sources
matched with KOBs have a $J$ band $\chi^2>2$, we concluded that the higher W1 and W2
excess occurrence rate was related to the poorer 2MASS source extraction. As with the
WISE W1 source extraction we attribute this correlation to confusion. We therefore only
use 2MASS data when the $\chi^2$ from source extraction for all three bands is less than
2.

We exclude 2MASS photometry for about 5,000 2MASS objects that have the E, F, X, or U
photometric quality flags. The first two flags represent the poorest quality photometry,
the third is for detections for which no brightness estimate could be made, and the last
is for upper limits.

We also used early SED fitting runs to assess the quality of the photometry in each
band. We found that Tycho 2 photometry is poorly suited to the task at hand. While the
measurements appear accurate, only a few thousand KOBs are brighter than about 11th
magnitude \citep{2010ApJ...713L.109B}, and objects fainter than this have large Tycho 2
uncertainties \citep[see][]{2000A&A...355L..27H}. Because most KOBs are near or beyond
the Tycho 2 magnitude limits, their precision is poor and we did not use this
photometry.

Finally, we found that the KIC $u$ photometry was commonly offset below the photospheric
models. Given that only 1092 stars have $u$ photometry we also excluded this band from
the fitting.

\subsubsection{W2 absorption}\label{sss:abs}

\begin{figure}
  \begin{center}
    \hspace{-0.5cm} \includegraphics[width=0.5\textwidth]{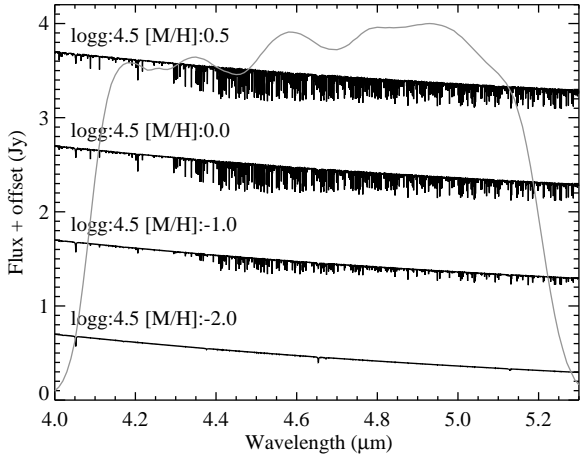}
    \caption{Metallicity-dependent absorption in the W2 band for $\log g = 4.5$ and a
      range of [M/H] for a Solar-type star. The W2 bandpass is also shown (arbitrary
      units, grey line).}\label{fig:abs}
  \end{center}
\end{figure}

The W2 band lies on top of the fundamental CO bandhead, whose depth varies strongly with
metallicity for the bulk of our sample. This effect is shown in Figure \ref{fig:abs},
where the lines of model spectra within the W2 bandpass (grey line) become deeper with
increasing metallicity. In this plot the flux at the W2 isophotal wavelength varies by
10\% between [M/H] of -2 to +0.5.

As the metallicity increases, the spectrum changes and the colour correction that should
be applied to the catalogue (sometimes called ``quoted'') flux also increases. Commonly,
IR colour corrections are simply taken to be those for a blackbody at the stellar
effective temperature. With this approach the examples in Figure \ref{fig:abs} would all
have the same colour correction. Thus, inaccuracies in the derived metallicities would
lead to a (spurious) trend of larger W2 excesses for more metal rich stars. However, when
computed properly the colour correction increases with the level of absorption by a
similar amount \citep[see also][]{2010AJ....140.1868W}. That is, in the metallicity range
considered at 5800K, the actual fluxes vary by about 10\% but the quoted fluxes vary by
only 1\%. WISE is therefore not actually very sensitive to metallicity for Sun-like
stars. This sensitivity depends on effective temperature and is strongest for M
dwarfs. This conclusion is borne out by the analysis in \S \ref{ss:sel} below.

\subsubsection{Final photospheric models}\label{sss:final}

The final SED models are generated based on the conclusions of this section. These were
computed for all but five of the WISE matches (that have no reliable photometry and are
identified in the KIC as galaxies). With the photospheric predictions in the W2-4 bands
made based on the optical and near-IR photometry, we now look for excesses.

\subsection{Selecting stars with excesses}\label{ss:sel}

With a complete set of SEDs, the task of finding excesses is in principle very simple;
stars with excesses greater than a sensible significance threshold are selected. In
practise this step is complicated by several factors. Atmospheric models for M dwarfs are
known to overestimate the stellar flux in the Rayleigh-Jeans regime
\citep[e.g.][]{2009ApJ...705...89L}. Some stars have poor fits and many of these result
in excesses that are not real and should be excluded. The sample also contains many
giants, whose excesses may be attributed to mass loss rather than a debris disk.

\begin{figure}
  \begin{center}
    \hspace{-0.5cm} \includegraphics[width=0.5\textwidth]{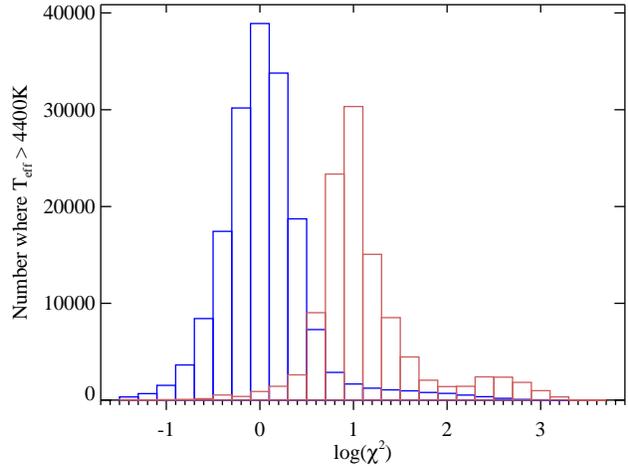}
    \caption{SED reduced $\chi^2$ for stars hotter (filled bars) and cooler (empty bars)
      than 4400K (number multiplied by 14). Cooler stars clearly have poorer fits, due
      either to missing opacity or incorrect bandpasses.}\label{fig:sedchi}
  \end{center}
\end{figure}

In our sample, there is a clear transition in the quality of the SED fits around
4400K. Stars cooler than this value have consistently poorer fits than those that are
hotter. Such a trend may be due to missing opacity in the stellar atmospheres, but could
also be caused by poor filter characterisation. Figure \ref{fig:sedchi} shows $\chi^2$
histograms for the SED fitting, where $\chi^2$ is the reduced sum of squared differences
between the photometry and the photospheric model. When split by effective temperature at
4400K there is a clear difference between the two sets, so we apply a different cut in
$\chi^2$ for each; for stars hotter than 4400K we keep stars with $\chi^2<10$, for those
cooler than 4400K we keep stars with $\chi^2<100$. We do not simply ignore these cool
stars, because the photospheric predictions are generally reliable (though not always, as
we find in \S \ref{ss:w12}).

\begin{figure*}
  \begin{center}
    \hspace{-0.5cm} \includegraphics[width=1\textwidth]{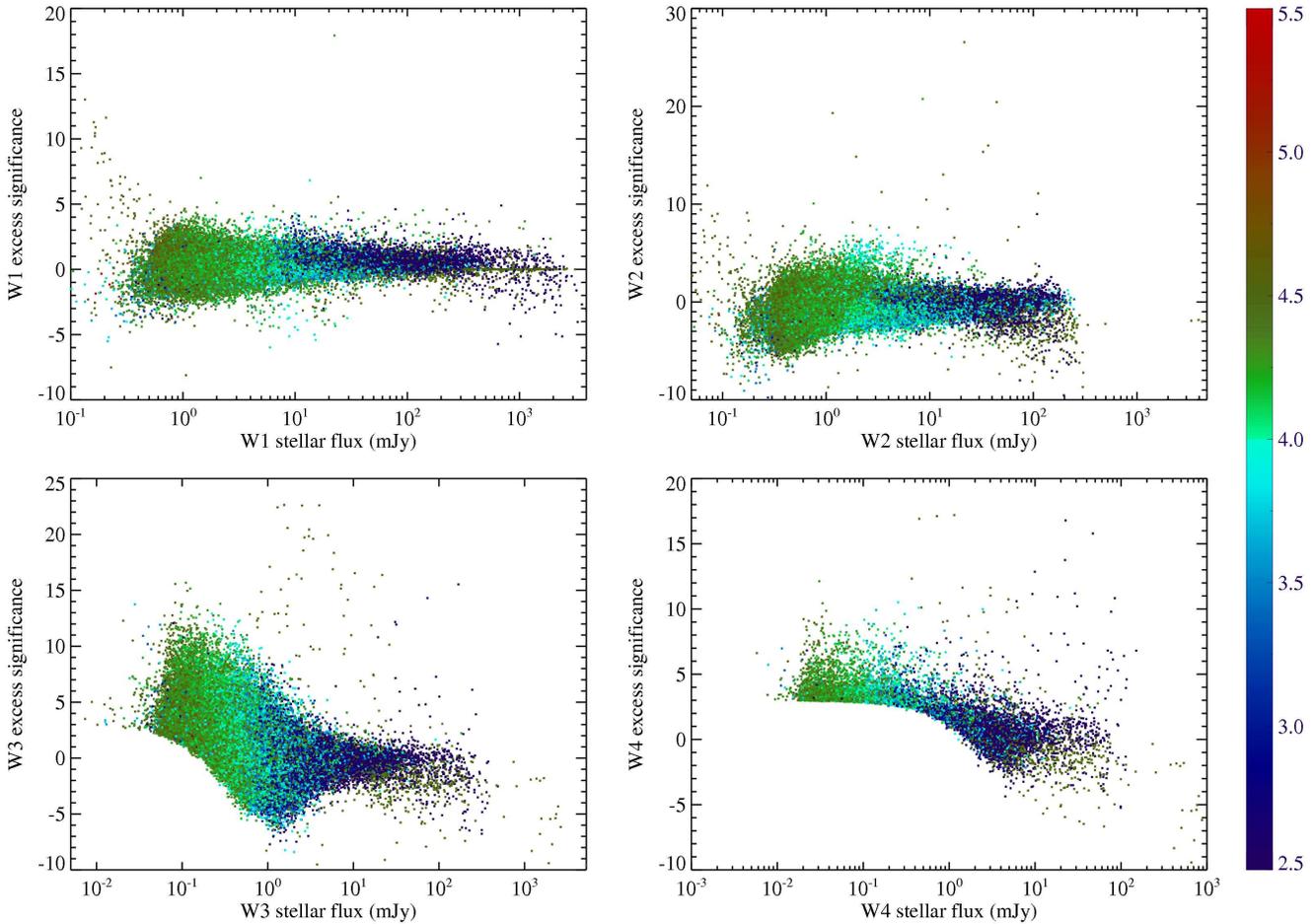}
    \caption{Excess significance vs. predicted photospheric flux in the W1-4 bands. The
      colour scale is $\log g$. Any object with significance greater than 3-4 plausibly
      has a real excess.}\label{fig:xs}
  \end{center}
\end{figure*}

Having made this cut in the quality of the SED fits, Figure \ref{fig:xs} shows the excess
significance for bands W1-4. Though we have included W1 in the photospheric fitting, if
the photometry in the shorter wavelength bands is of high quality then W1 will still show
an excess if present. The dot colours indicate the gravity derived for the KIC (and used
by us with the offset noted above), and show that some W3-4 excesses are present around
stars with lower gravities (i.e. bluer dots around brighter stars with excess
significance above 3-4). We therefore remove giants using the criteria of
\citet{2011AJ....141..108C}, where a star with $T_{\rm eff}>6000$ is assumed to be a
giant if $\log g<3.5$ and a star with $T_{\rm eff}<4250$ is assumed to be a giant if
$\log g<4.0$, with a linear transition for intermediate temperatures. Because we have
adjusted the gravity of giants as derived in the KIC, we likewise shift their giant
criterion down accordingly so our criterion selects the same stars.

A feature present in Figure \ref{fig:xs} for the W3-4 bands is a flux-dependent cut-off
below about 0.5 and 5mJy respectively. This cut-off is simply the WISE sensitivity limit,
which shows that the faintest stars can only be detected if the W3-4 flux is greater than
the photosphere (due to statistical variation or real excess emission).

\begin{figure}
  \begin{center}
    \hspace{-0.cm} \includegraphics[width=0.5\textwidth]{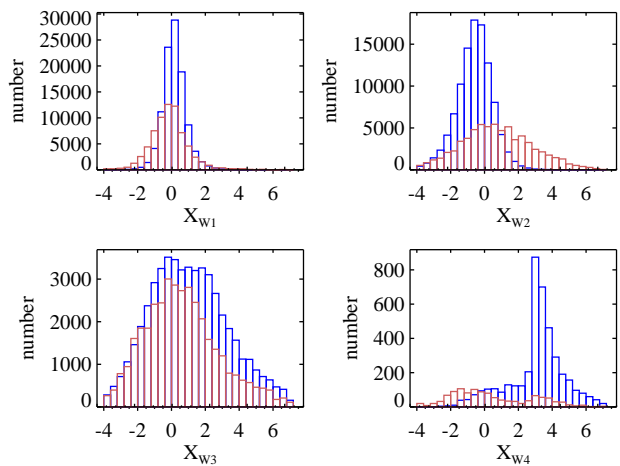}
    \caption{Excess significance histograms for the W1-4 bands for stars hotter (filled
      bars) and cooler (empty bars, number multiplied by 15, 15, 10, and 1 respectively)
      than 4400K.}\label{fig:sig}
  \end{center}
\end{figure}

Having removed poor SED fits and giants, the remaining task in identifying excesses is to
set the threshold level of significance. Normally this level would be around 3-4 if the
uncertainties are estimated appropriately, but there are a very large number of W3
excesses above this level. Figure \ref{fig:sig} shows the significance distributions, and
W3 clearly has many excesses that would be considered significant (as does W4, but to a
lesser degree). These cannot be debris disks, because the excesses that WISE can detect
around \emph{Kepler} stars are rare. The distribution should therefore appear largely
Gaussian with a dispersion of unity, with only a few objects at higher positive
significance. Aside from being affected by sensitivity limits as seen for W3-4, the
negative side of the histogram should be Gaussian (i.e. negative excesses cannot arise,
even if positive excesses arise due to true astrophysical phenomena), and the extent can
be used to estimate a reasonable significance threshold. Because the histograms do not
show negative excesses below a significance of $-4$, we set the threshold at $+4$ for
W1-4, and address the origin of the large number of excesses below.

We make an exception to this threshold for W2 excesses around cooler stars. The
significance distribution for W2 is much wider than for hotter stars and skewed to larger
values (Fig. \ref{fig:sig}). This difference presumably arises due to greater absorption
in the W2 band (see \S \ref{sss:abs}). Plotting the significance against metallicity
indeed shows a strong correlation, which could either be a sign that W2 excesses around M
stars are strongly correlated with metallicity or that the metallicity of these stars in
the KIC is too high (i.e. the absorption in the model is stronger than in reality). Given
that debris disks around nearby M dwarfs appear to be very rare
\citep{2006A&A...460..733L,2007ApJ...667..527G,2009A&A...506.1455L} and show no such
trend, the latter is the more sensible conclusion and we set the significance threshold
at 7.

\section{Interpretation of excesses}\label{s:interp}

With our chosen significance criteria, there are 7,965 disk candidates. There are 79, 95,
7480, and 1093 excesses in bands W1-4 respectively. These excesses correspond to an
occurrence rate of about 4\%. Since about 4\% of nearby Sun-like stars have 24$\mu$m
excesses \citep[e.g.][]{2008ApJ...674.1086T} from calibration limited observations (flux
ratios $\gtrsim$1.05), the finding of a similar rate from much less sensitive WISE
observations (see Fig. \ref{fig:sens}) indicates that unless the stars observed by
\emph{Kepler} are somehow unique, most of the excesses cannot be due to debris. We
therefore take a closer look at the origins of these excesses in the next two
subsections. In what follows, we group stars into three effective temperature bins;
``M-type'' ($T_{\rm eff}<4400$K), ``FGK-type'' or ``Sun-like'' ($4400 < T_{\rm eff} <
7000$K), and ``A-type'' ($7000 < T_{\rm eff}<10,000$K). Only ten excesses are found for
stars hotter than 10,000K, all in W3, and none survive the following analysis.

\subsection{W1-2 excesses: poor photospheric predictions}\label{ss:w12}

\begin{figure}
  \begin{center}
    \hspace{-0.5cm} \includegraphics[width=0.5\textwidth]{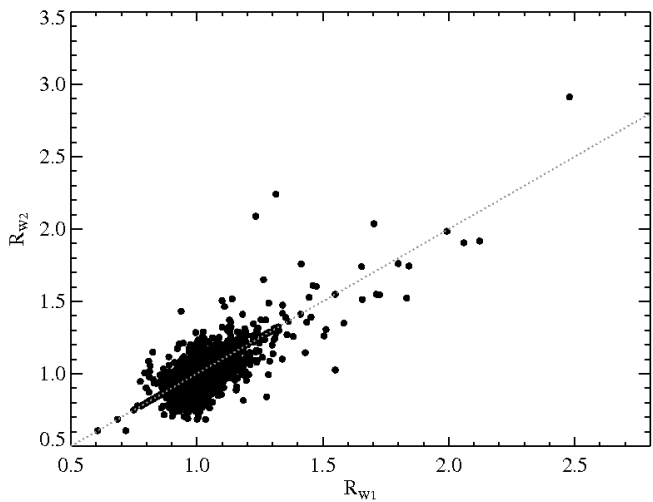}
    \caption{Flux ratios in the W1 and W2 bands compared for all stars with excesses (the
      dotted line is $y=x$, not a fit to the data). The strong correlation is not
      expected for dust emission and is due to poor photospheric
      predictions.}\label{fig:w1w2}
  \end{center}
\end{figure}

A handful of targets show W1-2 excesses, but Figure \ref{fig:w1w2} argues that they are
probably not due to circumstellar debris. Plotting the flux ratios in the W1 and W2 bands
shows that these quantities are correlated, with a slope of approximately unity. The
excess can therefore be accounted for by shifting the stellar spectrum upward.

Inspection shows that the objects with the largest ($\gtrsim1.5$) flux ratios in W1-2 are
the result of failed photospheric fits, where the optical photometry is at odds with the
WISE photometry. In these cases the stellar temperature is generally below 4400K, and
were not cut due to the relaxed photospheric fit $\chi^2$ for these objects. These
objects typically have no temperature in the KIC, meaning that no reasonable fit could be
found there either.

Some objects have smaller flux ratios in W1-2 that are also significant, but these ratios
remain well correlated. While some are still due to poor photospheric predictions,
another explanation is that the W1-2 photometry includes two stars. The excess flux in
the W1-2 bands could be caused by the emission from a cooler star that lies within the
instrumental PSFs of all photometry (and may or may not be associated with the
\emph{Kepler} star in question). It is also possible that the higher resolution 2MASS and
optical photometry used to predict the photosphere measured flux from one of a pair of
stars, while WISE measured flux from both. Such a situation can also lead to
identification of an excess where there is none.\footnote{An example is HIP 13642,
  identified by \citet{2010ApJ...710L..26K} as an excess because the MIPS observation
  includes two stars, while the 2MASS observation used to predict the photosphere
  resolves the pair.}

Based on the strong correlation between the flux ratios in W1 and W2, we conclude that
while some excesses are likely real in that the spectrum departs from our model of a
single stellar photosphere, it is unlikely that any are excesses are due to circumstellar
dust.

\subsection{W3-4 excesses: disks or background?}\label{ss:w34}

While we have taken care to remove spurious detections (see \S \ref{sss:susp}), the WISE
sensitivity and resolution and the very large sample size mean that extra-Galactic
contamination due to chance alignments, even at very low levels, could contribute to, or
even be the cause of, the W3-4 excess population. Further, the low Galactic latitude of
the \emph{Kepler} field means that IR flux levels from dust within our Galaxy can be
significant (see Fig. \ref{fig:w1cov}).

\subsubsection{Galactic background contamination}\label{ss:bg}

\begin{figure}
  \begin{center}
    \hspace{-0.3cm} \includegraphics[width=0.5\textwidth]{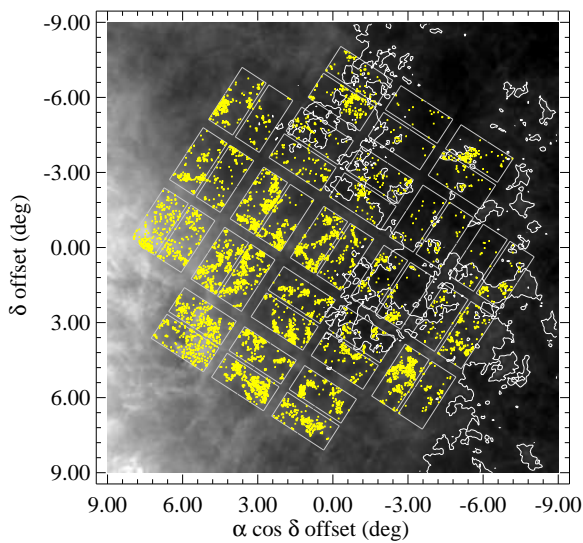}
    \caption{Clumping of stars with W3 excesses (yellow dots) indicating that the
      excesses are due to the high background level. The 5MJy/sr cut based on the IRAS
      100$\mu$m background image is shown by the white contours.}\label{fig:w3xs}
  \end{center}
\end{figure}

\begin{figure*}
  \begin{center}
    \hspace{-0.5cm} \includegraphics[width=0.5\textwidth]{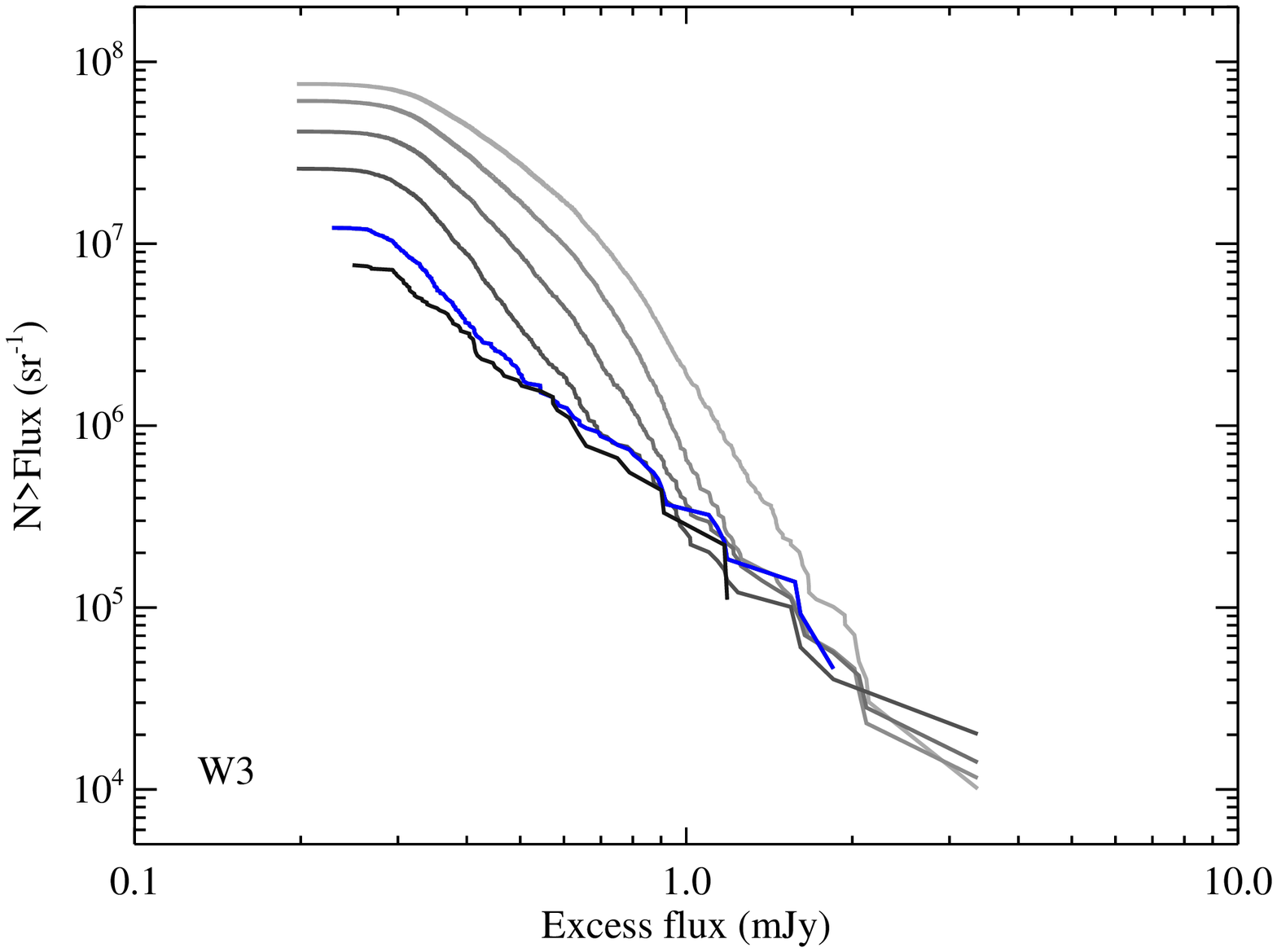}
    \includegraphics[width=0.5\textwidth]{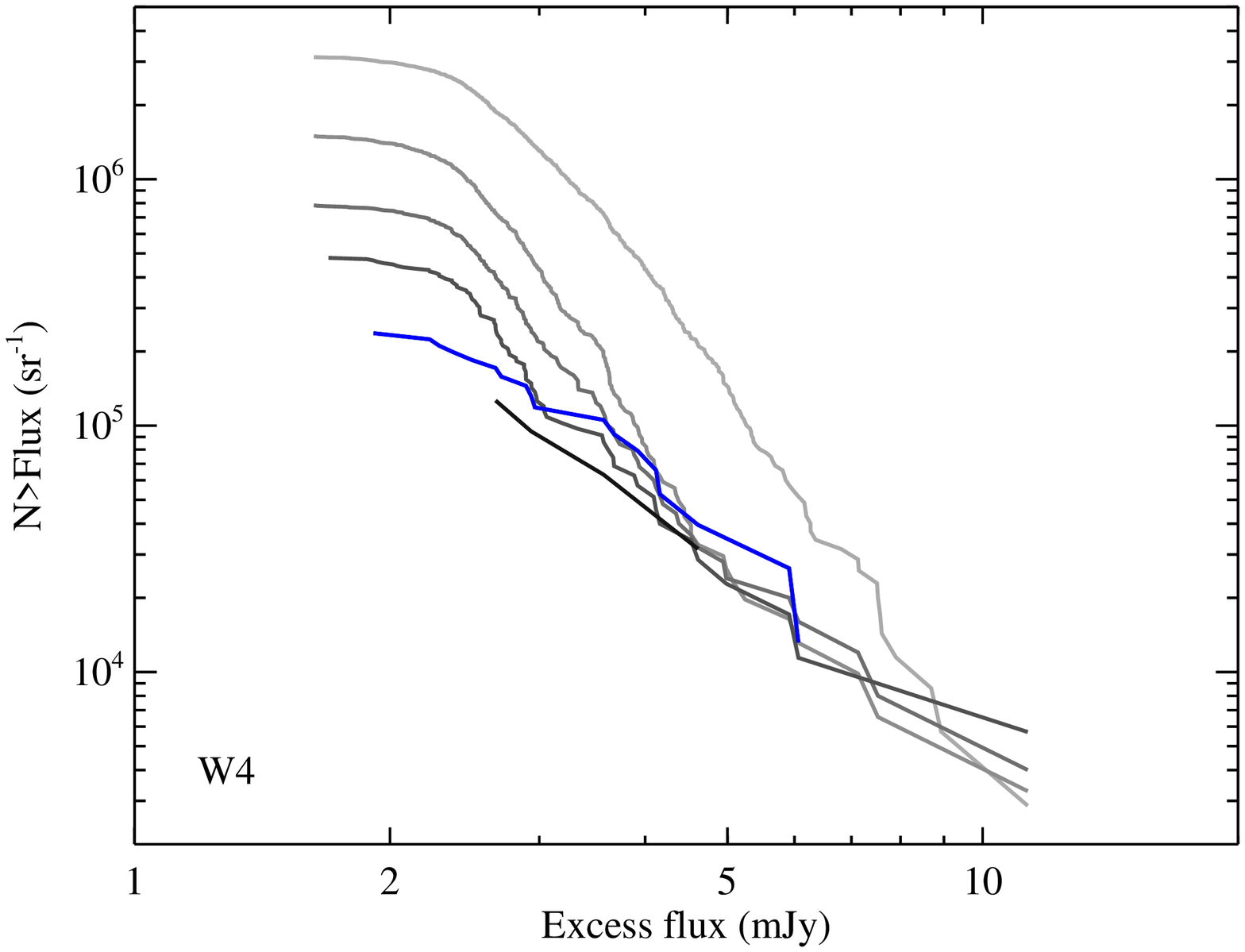}
    \caption{Cumulative source counts in W3 (left panel) and W4 (right panel). The solid
      lines show the counts for excesses as the cut in IRAS 100$\mu$m background flux is
      increased. The lines show no cut (top line), and then levels of 15, 10, 7, 5, and
      4MJy/sr. As the cut in IRAS background increases the excess counts approach a fixed
      level, indicating that the remaining excesses are not caused by Galactic background
      emission. The blue line shows our adopted cut level of 5MJy/sr.}\label{fig:bgcut}
  \end{center}
\end{figure*}

The hypothesis that the Galactic background level is the cause of the very large number
of W3 excesses can be tested by simply plotting their locations on the sky, shown as dots
in Figure \ref{fig:w3xs}. The excesses clearly reside in clumps, and appear more frequent
closer to the Galactic plane. Therefore, the bulk of the W3 excesses are likely spurious.

To remove these false excesses in a way unbiased for or against the presence of excess
emission therefore requires ignoring excesses in the highest background regions. Ideally
this cut would be made based on the WISE catalogue itself. In general however, the
background (\texttt{w3sky} column in the catalogue) is smooth, and shows no relation to
the clumpiness seen for excesses except very near to the Galactic plane. This smoothness
is perhaps a result of the dynamic WISE calibration, which attempts to remove temporal
instrumental
variations.\footnote{\href{http://wise2.ipac.caltech.edu/docs/release/allsky/expsup/sec4\_4a.html}{http://wise2.ipac.caltech.edu/docs/release/allsky/expsup/sec4\_4a.html}}
We found that instead the 100$\mu$m IRAS IRIS map is a very good indicator of the
background level, which we use to exclude sources below.

Figure \ref{fig:bgcut} shows how the cumulative number of excesses changes as a function
of the IRAS background level. To make the plots comparable with those below, we scale the
number of excesses by dividing by the area covered by the WISE observations. We take the
observed area for a single star to be that enclosed by a circle whose diameter is the
WISE point spread function (PSF) full-width at half-maximum (FWHM, 6\farcs5 and 12'' for
W3-4 respectively). These areas are multiplied by the number of non-giant stars with
satisfactory SED fits that were observed and lie in regions below the given background
level (and for which photometry was not removed for any of the reasons in \S
\ref{sss:susp}).

In each plot the highest line shows the full set of excess counts. The lower lines show
how the excess counts decrease as an increasing cut in the IRAS 100$\mu$m background
level is made. Once the cut level reaches about 5MJy/sr the excess counts stop
decreasing, indicating that the excesses that are due to the high background level have
been removed. Higher cut levels do not decrease the distributions further and simply
result in fewer remaining excesses.

The region where the IRAS background level is lower than 5MJy/sr is shown in Figure
\ref{fig:w3xs}. The contours mark out and avoid regions where excesses clump together
well. Based on this approach, we conclude that 5MJy/sr is a reasonable cut level to avoid
contamination from the high Galactic background level. Of the initial 7,965 disk
candidates, 271 remain after this cut.

\subsubsection{Extra-Galactic counts}\label{ss:counts}

\begin{figure*}
  \begin{center}
    \hspace{-0.5cm} \includegraphics[width=0.5\textwidth]{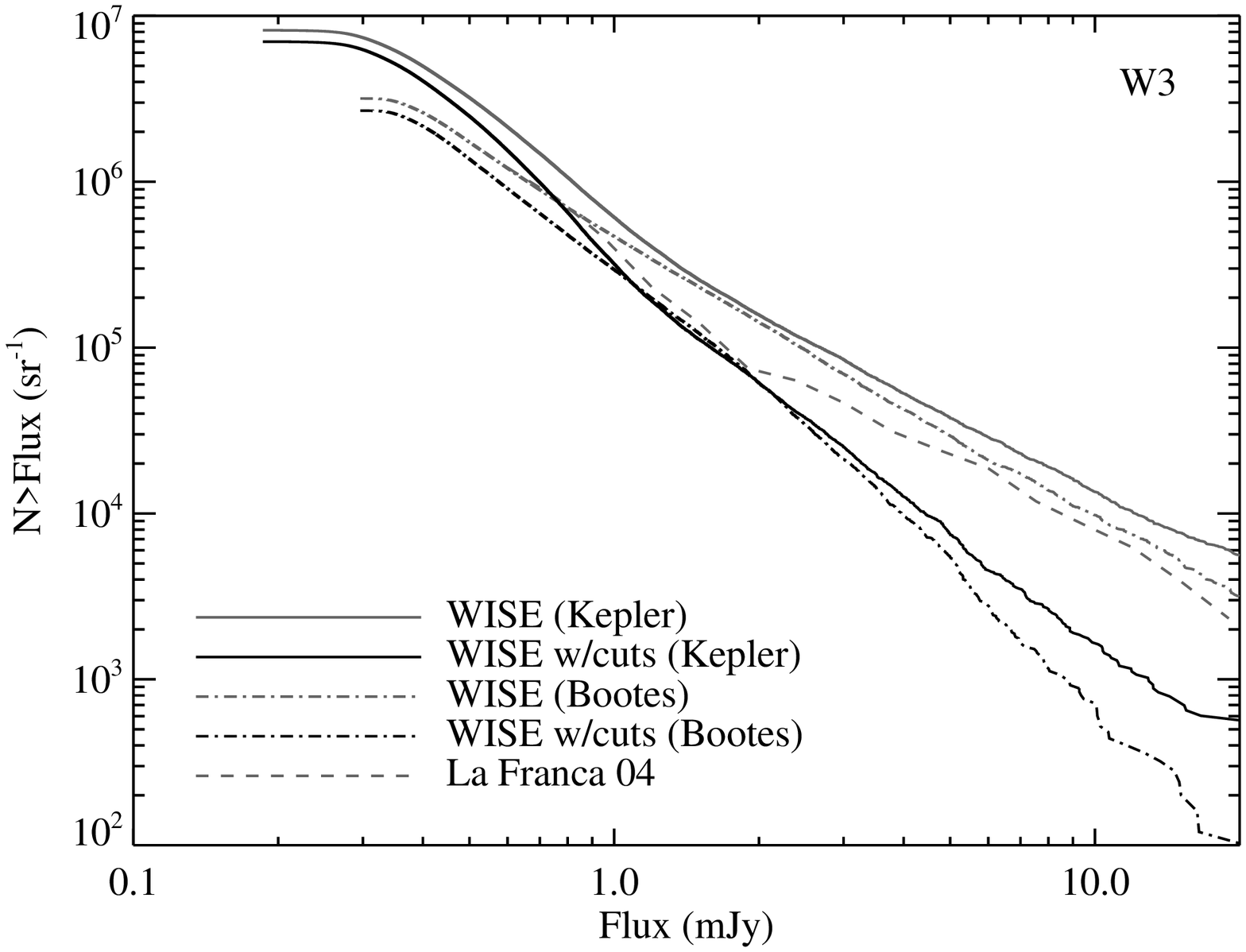}
    \includegraphics[width=0.5\textwidth]{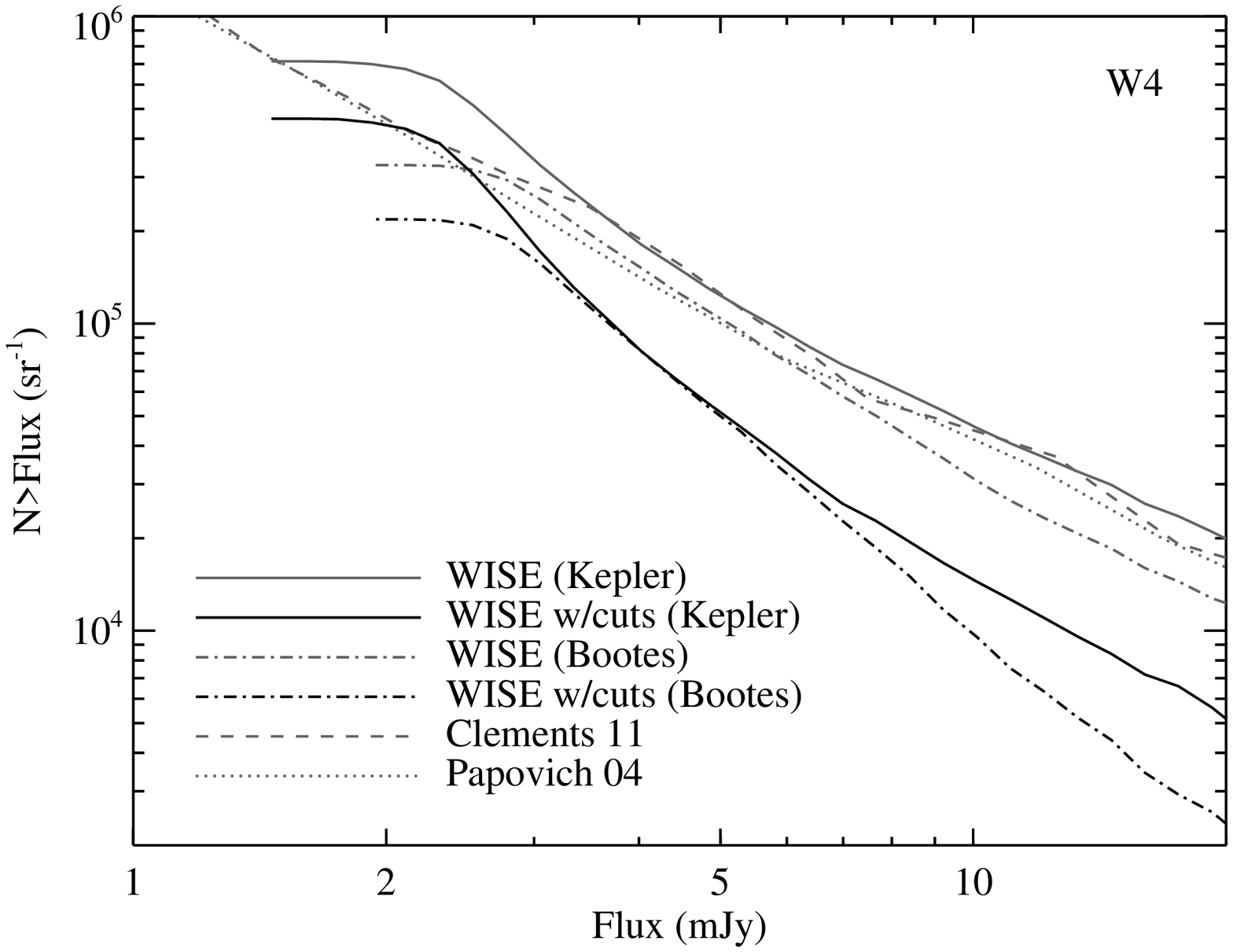}
    \caption{Comparison of cumulative galaxy source counts in W3 (left panel) and W4
      (right panel). The grey lines show galaxy counts from ISOCAM at 15$\mu$m
      \citep[left panel,][]{2004AJ....127.3075L} and \emph{Spitzer} MIPS at 24$\mu$m
      \citep[right panel,][]{2004ApJS..154...70P}, and from WISE in two different fields
      (see text). The black lines show the same WISE fields, but with the additional cuts
      outlined in \S \ref{sss:susp} (i.e. the same cuts as were applied in our search for
      excesses around \emph{Kepler} stars).}\label{fig:gal}
  \end{center}
\end{figure*}

The remaining 271 excesses are generally real in the sense that they arise from
point-like flux above the photospheric emission at the location of the \emph{Kepler}
stars. However, we now test whether these could arise from chance alignments with
background galaxies. To estimate the number of excesses expected from extra-Galactic
contamination we therefore first derive galaxy counts specific to our sample. The galaxy
counts are derived by counting the number of sources above a given flux at a given
wavelength after the contribution of Galactic stars has been removed.

Because galaxy counts may be subject to cosmic variance, and could appear to be different
in the \emph{Kepler} field due to stellar crowding and a relatively high background level
near the Galactic plane, we show the results from several different fields and surveys in
Figure \ref{fig:gal}. For comparison with WISE W3 we show 15$\mu$m ISO results
\citep{2004AJ....127.3075L} and for W4 we show 24$\mu$m \emph{Spitzer} MIPS results
\citep{2004ApJS..154...70P,2011MNRAS.411..373C}. We compare these with counts from two
fields we extracted from the WISE catalogue. The first is a box in the \emph{Kepler}
field between 286-296$^\circ$ right ascension and 40-50$^\circ$ declination (71 square
degrees). The second is a ``random'' box farther away from the Galactic plane in
Bo\"otes, between 210-220$^\circ$ right ascension and 30-40$^\circ$ declination (82
square degrees, at a Galactic latitude of about 70$^\circ$).

For our analysis of the WISE data we require A, B, or C quality photometry
(\texttt{ph\_qual}), and ${\rm S/N} > 4$, and remove the stellar contribution by keeping
sources with ${\rm W1} - {\rm W3,4} > 1.2$ \citep[see][]{2011ApJ...735..112J}. These
source counts are shown in Figure \ref{fig:gal} in grey, and follow the ISO and
\emph{Spitzer} counts well. The agreement suggests that cosmic variance is not
significant for these fields (i.e. the distribution of background galaxies is similar in
the \emph{Kepler} field to elsewhere). The WISE counts are similar for both fields,
though the \emph{Kepler} field shows somewhat increased counts at the lowest flux levels.

We then add the cuts outlined in \S \ref{sss:susp}, which were made with the intention of
minimising galaxy contamination, for which the results are shown as black lines. The
black lines lie below the grey ones, indicating that the extra cuts do indeed remove some
galaxies. The cuts are more effective in W4 with an overall decrease, while at W3 the
cuts are only effective for brighter galaxies. Because we want to quantify the
extra-Galactic contribution to our excesses, we use the black line from the \emph{Kepler}
field as the expected level of galaxy contamination.

\subsubsection{Extra-Galactic contamination}\label{ss:bgal}

\begin{figure*}
  \begin{center}
    \hspace{-0.5cm} \includegraphics[width=0.5\textwidth]{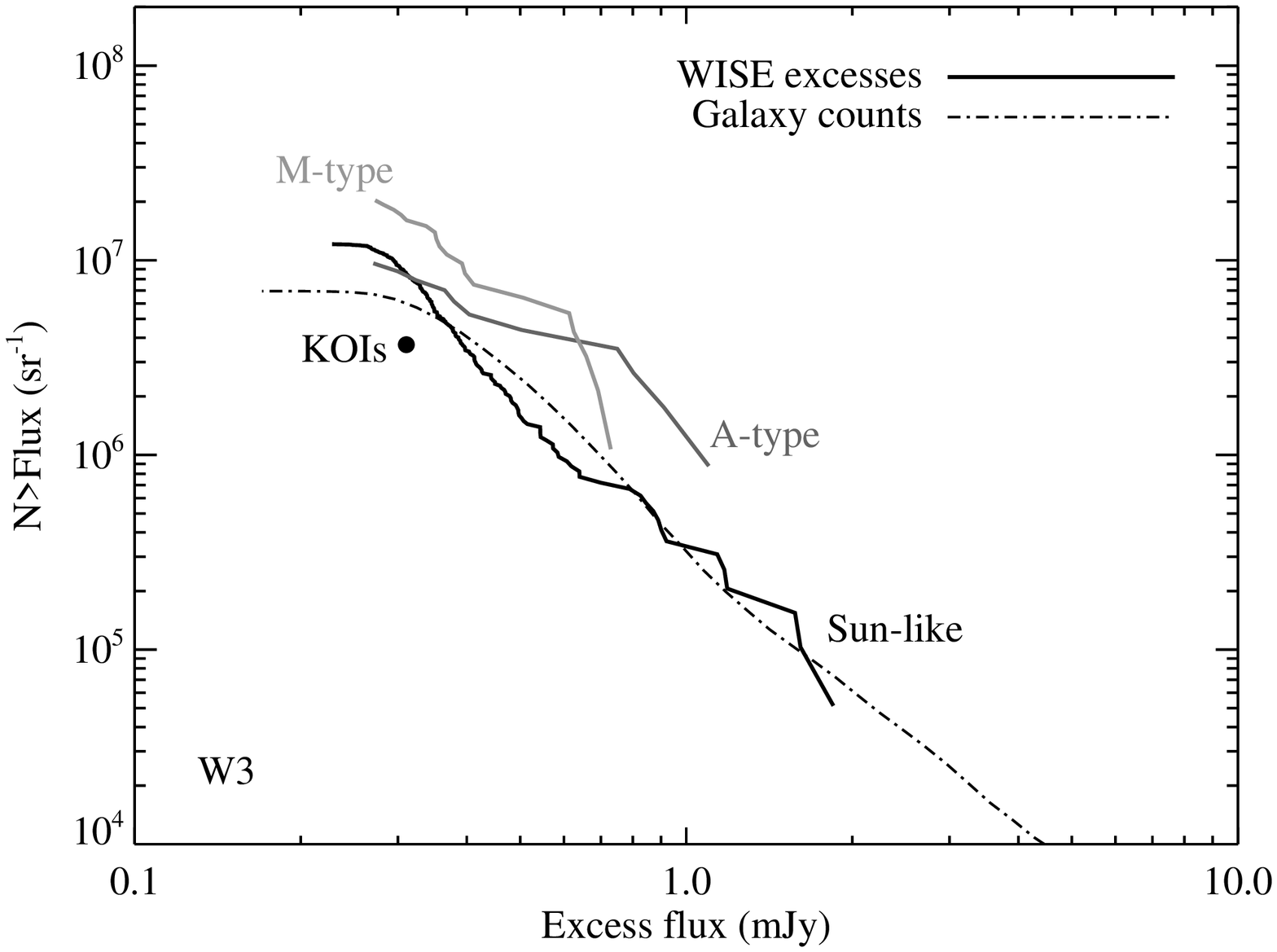}
    \includegraphics[width=0.5\textwidth]{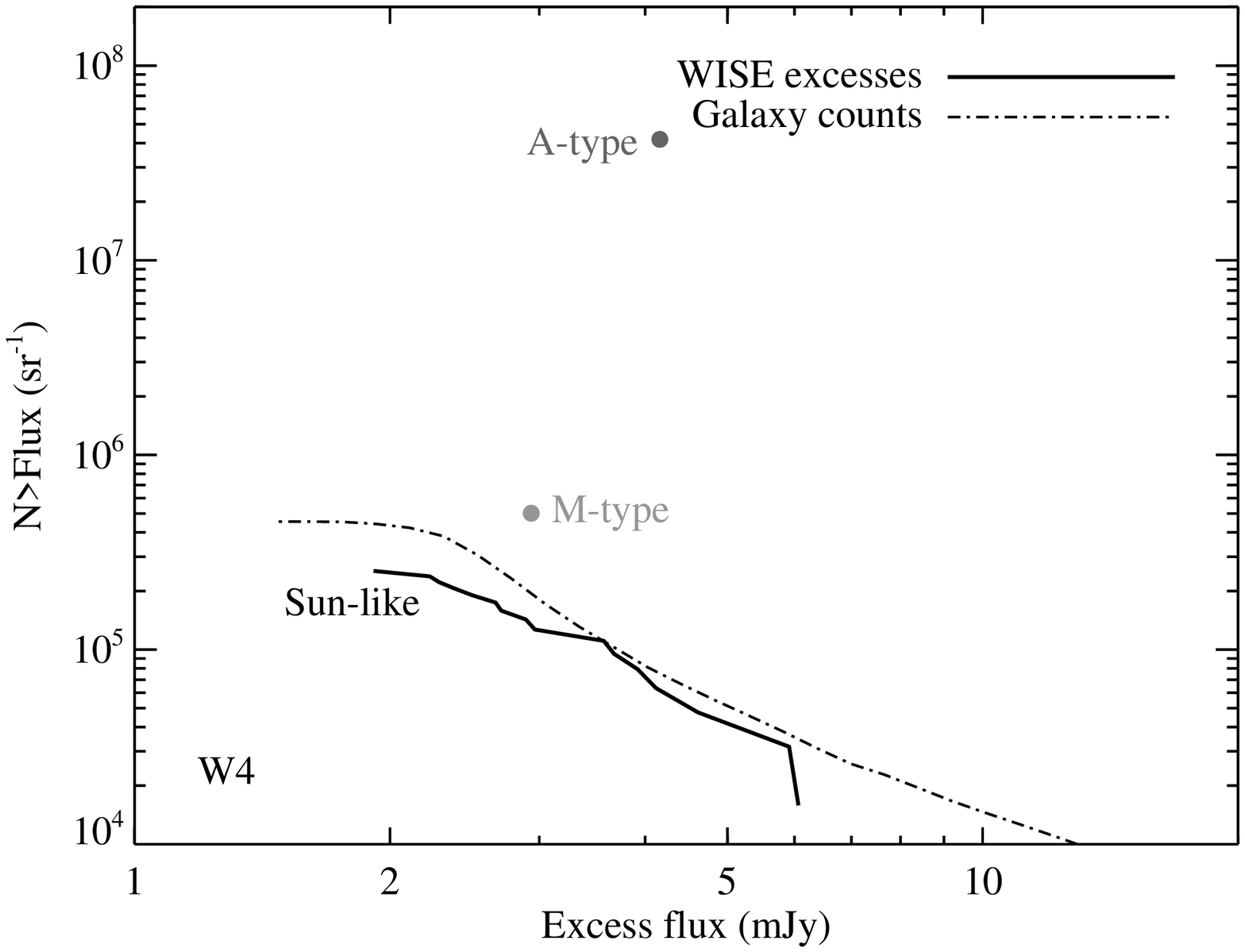}
    \caption{Cumulative source counts in W3 (left panel) and W4 (right panel). The solid
      lines shows the counts for excesses, split into M-type, Sun-like and A-type (there
      are no W4 excesses around M-types). The dot-dashed line shows our WISE counts in
      the \emph{Kepler} field (solid lines in Fig. \ref{fig:gal}). The dot is the single
      planet host candidate found to have an excess (KOI 861). All excesses lie near the
      level expected from background galaxies, with the exception of a single W4 excess
      around a nearby A-type stars.}\label{fig:counts}
  \end{center}
\end{figure*}

\begin{table*}
  \caption{The 271 \emph{Kepler} stars with WISE 3-4 excesses (full table in the appendix
    of this arXiv version). Columns are: KIC identifier, predicted \emph{Kepler} ($K_p$)
    magnitude from the KIC, Quarters the star
    was observed in (up to 6), fitted effective temperature, W3-4 flux ratio and excess
    significance (where $X_{\rm W3,4} \ge 4$). The note column notes the single KOI,
    and potential planet hosts from \citet{2012arXiv1201.1048T} (``T12'').}
  \begin{tabular}{lrrrrrrrl}\label{tab:xs}
    KIC & $K_p$ & Quarters & $T_{\rm eff}$ & $R_{\rm W3}$ & $X_{\rm W3}$ & $R_{\rm W4}$
    & $X_{\rm W4}$ & Notes \\
    \hline
    5866211&15.19&456&6585&4.2&5.0&&&\\
5866341&15.06&123456&6296&4.9&6.7&&&\\
5866415&15.33&123&6029&4.7&4.4&&&\\
6198278&14.86&123456&5436&2.8&4.1&&&\\
6346886&14.96&12346&5869&3.9&4.6&&&\\
6431431&14.87&123456&8147&5.9&7.8&&&\\
6503763&15.78&12346&5275&4.3&4.1&&&\\
6515382&13.29&123456&6265&1.7&4.2&&&\\
6516101&13.88&123456&6062&2.3&6.1&&&\\
6599949&15.42&123456&5773&4.0&4.1&&&\\
6676683&14.58&123456&6356&4.1&6.2&&&\\
6685526&15.00&123456&5103&2.7&4.1&&&KOI 861,T12\\

  \end{tabular}
\end{table*}

We now proceed with the remaining excesses where the IRAS 100$\mu$m background is lower
than 5MJy/sr (listed in Table \ref{tab:xs}). The comparison of these excesses (again
expressed as counts per sky area) and the galaxy counts derived above is shown in Figure
\ref{fig:counts}, where we have now separated the excesses by spectral type. In these
plots, the galaxy counts do not move. The excess counts from disks depend on their
occurrence rate and the distance to the stellar sample. Naturally, a higher disk fraction
would move the excess counts upward on this plot, away from the galaxy counts, and the
chance of an individual excess being due to a background galaxy would be lower. For a
fixed excess distribution, samples of stars that are on average fainter and brighter
(i.e. farther and nearer), move the excess count lines left and right respectively. Thus,
the excess counts from disks for samples of brighter stars lie higher above the galaxy
counts than samples of fainter stars, and again the likelihood of contamination is
lower. This advantage arises because for fixed disk to star flux ratio (i.e. fixed disk
properties), the absolute flux from a debris disk around a bright star is more than that
from a faint star. At brighter flux levels the number of galaxies per unit sky area is
smaller, so the likelihood of confusion lower. Finally, higher instrument resolution
means less chance of confusion with a background galaxy because the area surveyed per
star is smaller. Therefore, the same population of excesses observed with a larger
telescope would also be further above the galaxy counts and more robust to confusion.

In Figure \ref{fig:counts}, we make one additional cut to the number of stars that count
towards the total area observed, by only including stars whose photospheres are equal to
or brighter than that of the faintest star found to have an excesses. This photospheric
flux cut makes use of the fact noted above, that brighter stars are more robust to
confusion (assuming that the presence or otherwise of a disk is independent of stellar
brightness for fixed spectral type). This cut has little effect for most excesses because
the faintest star with an excess is near the limit for all stars. However, it is
effective for the W4 excess associated with an A-type star because this star is brighter
than the bulk of the sample. In W3 the total number of non-giant stars that survive the
cut in background level is 1198, 24916, and 1462 for M, FGK, and A stars respectively. In
W4 the numbers are 750, 23742 and 10 for M, FGK and A-types. The very small number of
A-type stars as bright or brighter than the one with an excess shows why the cut in
photospheric flux is useful.

For W3 (left panel of Fig. \ref{fig:counts}) the excess counts for 19 M-type, 235
Sun-like and 11 A-type stars lie very close to the counts expected from background
galaxies. Thus, not many, if any, of the excesses appear attributable to debris disk
emission. The disk occurrence rate is insufficient to allow detection of debris disks
that are robust to galaxy confusion (e.g. have a less than 1/10 chance of being a
galaxy). The single \emph{Kepler} planet host candidate (\emph{Kepler} Object of
Interest, or KOI) found to have an excess \citep[(KOI 861 a.k.a. KIC
6685526,][]{2011ApJ...736...19B,2012arXiv1202.5852B} is shown as a single point, as one
of 348 KOIs that survive the IRAS background cut. It lies very close to the A, FGK, and
M-type sample counts, so is equally likely to be confused. Because the excesses are
heavily contaminated, the excess distributions represent an upper limit on the
distribution of W3 excesses (we return to these limits in \S\S \ref{ss:stat12} and
\ref{s:disc}).

For W4 (right panel of Fig. \ref{fig:counts}) the 16 Sun-like excess counts lie sightly
below the galaxy counts, the single M-type excess slightly above, while the single A-type
excess lies well above. Because galaxy counts are independent of stellar spectral type,
there should be no difference between the contamination level for M-type, Sun-like and
A-type stars. Therefore, the 22$\mu$m A-type excess, which has a moderate 22$\mu$m flux
ratio of 1.63, is very likely due to debris disk emission. Because the difference between
the A-type excess counts and the WISE galaxy counts is about a factor of one hundred,
there is about a 1/100 chance that this A-star excess is a galaxy. It is likely that all
Sun-like W4 excesses and the single M-type excess can be explained as galaxy confusion,
so again the disk occurrence rate is too low to allow robust disk detection and the
excess counts represent an upper limit.

It is perhaps surprising that the excess and galaxy counts in Figure \ref{fig:counts}
agree as well as they do. The galaxy counts were derived from all sources that met
certain criteria within a specific patch of sky with the assumption that confusion only
happens within the WISE PSF FWHM, while the excess counts were the result of the SED
fitting method using WISE photometry at positions of known stars. We applied a cut in the
background level to remove spurious excesses, but no such cut was required for the galaxy
counts. The extra-Galactic counts in the \emph{Kepler} field agree well with those for
the Bo\"otes field, where the IRAS 100$\mu$m background level never reaches more than
about 3MJy/sr (i.e. is always below our cut in background level), so the extra-Galactic
counts in the \emph{Kepler} field are relatively unaffected by the background. Therefore,
there appears to be a preference for stars (which are almost always detected in W1-2) to
show a spurious W3-4 flux due to high background levels, while galaxies (which are
generally not detected in W1-2) do not. This difference may be attributed to the WISE
method of source extraction, which attempts to measure fluxes across all four bands if a
source is detected in at least one.

\subsection{Comparison with previous results}\label{ss:prev}

Our study is not the first to use WISE to look for warm emission from disks around
\emph{Kepler} stars
\citep{2012arXiv1203.0013R,2012ApJ...752...53L}. \citet{2012arXiv1203.0013R} found 13
candidate disk systems using the WISE preliminary release, 12 of which are observed by
\emph{Kepler} \citep[the other is WASP-46, a nearby system with a transiting
planet,][]{2012MNRAS.422.1988A}. However, they use an excess significance threshold of 2
(see eq. \ref{eq:chix}). At this level 2.3\% of systems are expected to have significant
excesses purely due to the fact that the uncertainties have a distribution (that is
assumed to be Gaussian). Therefore, of the 468 \emph{Kepler} planet host candidates they
considered, 11 should lie above this threshold. This number is similar to their 12 disk
candidates, so these candidates are consistent with being part of the expected
significance distribution if no stars have disks.

Three of their twelve disk candidates have significance higher than 3, but all lie in
regions where the 100$\mu$m background is higher than 5MJy/sr so are excluded from our
analysis because their excesses are likely due to the high background level (\S
\ref{ss:bg}).\footnote{KOI 469 has a very bright moving object (i.e. an asteroid or
  comet) visible in the WISE images at a separation of about 6 arcminutes, which may have
  affected the source extraction. Given the rarity of excesses, it seems more likely that
  the apparent excess is due to the presence of the bright object, rather than
  coincidental.}

In contrast to our conclusions, \citet{2012arXiv1203.0013R} find that background
contamination is negligible, with a $5 \times 10^{-5}$ chance of a galaxy brighter than
5mJy appearing within 10'' of a source at 24$\mu$m, using counts from
\citet{2004ApJS..154...70P}. However, these counts show about $10^5$ sr$^{-1}$ for
sources brighter than 5mJy (see Fig. \ref{fig:gal}), so a target area of 314 square
arcseconds (10'' radius) yields $10^5 \times 314 \times 2.35 \times 10^{-11} = 0.001$
probability of having a 5mJy background source within 10'' of a target. However, the WISE
beam is in fact smaller than 10'' radius, so using 6'' is more appropriate (see previous
subsection). Furthermore, removing WISE photometry that is flagged as extended decreases
the W4 counts (Fig. \ref{fig:gal}), so the \citet{2004ApJS..154...70P} counts
overestimate the confusion level that applies here by a factor of about two. Therefore,
0.06 spurious excesses are expected from 468 targets. This expectation is in line with
the two sources they report with W4 excesses, KIC~2853093 and KIC~6665695, since these
have W4 S/N of 2.2 and 2.5 respectively and as noted above we would not consider these
significant given the sample size.

All of their candidates have 12$\mu$m excesses, so should be compared with galaxy counts
at a similar wavelength (e.g. 15$\mu$m ISO counts). Based on Figure \ref{fig:gal}, about
$5 \times 10^6$ background galaxies per steradian are expected down to the detection
limit of about 3mJy, which for a target radius of 3.25'' (33 square arcseconds) yields an
expected contamination rate of $5 \times 10^6 \times 33 \times 2.35 \times 10^{-11}
=0.004$. Thus, about 2 spurious excesses among 468 targets is expected at this
wavelength. Of their W3 disk candidates, three have excesses more than 3$\sigma$
significant. However, KOI 1099 has W3-4 upper limits in the newer all-sky release, so the
expectation of two spurious excesses appears to be met. Both sources lie in the regions
we excluded due to the high background so the WISE photometry may still be spurious.

In a similar study, \citet{2012ApJ...752...53L} reported the discovery of excess emission
around eight \emph{Kepler} planet-host stars using the WISE Preliminary release. They
used a significance criterion of $5\sigma$, so their excesses should be astrophysical
(i.e. not statistical). There are only three candidates in common with
\citet{2012arXiv1203.0013R}. However, these are the three noted above with a significance
greater than $3\sigma$, so \citet{2012ApJ...752...53L} find the same candidates as
\citet{2012arXiv1203.0013R} with an additional five disk candidates. Of their eight, the
WISE W3 and W4 measurements of KOI 904 and KOI 1099 are upper limits in the newer all-sky
WISE catalogue, and KOIs 871, 943, 1020, and 1564 were rejected by
\citet{2012arXiv1203.0013R} after image inspection. The same two plausible disk
candidates remain, corresponding to the number estimated above to arise from
confusion. Though the details vary, these two studies are basically consistent if a
3-5$\sigma$ significance criterion is used and candidates are rejected based on the
images.

\subsection{Debris disk candidate}\label{ss:cand}

\begin{figure}
  \begin{center}
    \hspace{-0.5cm} \includegraphics[width=0.5\textwidth]{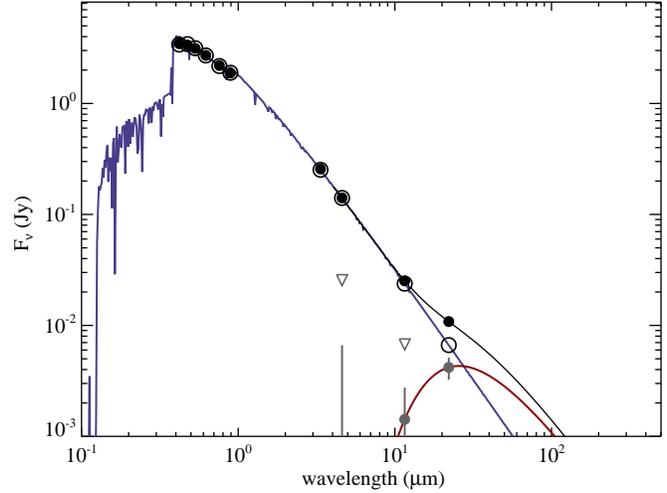}
    \caption{SED for the A-type W4 debris disk candidate KIC 7345479. The stellar
      spectrum is shown in blue, and the fitted blackbody in red. The sum of the two is
      shown in black. KIC and WISE photometry is shown as black dots, and synthetic
      photometry of the star in the same bands as open circles. Grey symbols show the
      star-subtracted fluxes, with a triangle above indicating the 4$\sigma$ upper limit
      if necessary.}\label{fig:sedw4}
  \end{center}
\end{figure}

We now briefly outline some properties of our most promising disk candidate, KIC 7345479
(with a \emph{Kepler} $K_p$ magnitude of 7.9), shown in Figure \ref{fig:sedw4}. Assuming
that this star is a dwarf yields a distance of 280pc, much closer than most \emph{Kepler}
stars. The SED shows the 9700K stellar spectrum, along with a simple blackbody fit to the
excess that includes the W2-4 photometry. With a measured flux of $10.8 \pm 0.9$mJy and a
photospheric flux of $6.6 \pm 0.2$mJy the W4 excess has a flux ratio $R_{\rm W4}=1.6$,
with significance $X_{\rm W4} = 4.5$. The disk temperature is constrained by the W3 upper
limit, so is cooler than about 200K, corresponding to a radial distance of greater than
15AU and lies beyond the region of \emph{Kepler} sensitivity to transiting planets. The
fractional luminosity for the blackbody model shown is $3.25 \times 10^{-5}$. As we show
in \S \ref{s:nearby} below, aside from the potential for planet discovery around the host
star, this disk is fairly unremarkable within the context of what is known about disks
around nearby A-stars.

\section{Nearby star comparison}\label{s:nearby}

There should be nothing particularly special about \emph{Kepler} stars compared to nearby
stars, so we compare our survey with 24$\mu$m results from two large unbiased
\emph{Spitzer} surveys of nearby stars. Because the results can only be interpreted
within the context of what was possible with each survey, we first compare the
sensitivity to disks for the \emph{Spitzer} surveys in the fractional luminosity
vs. temperature space introduced in Figure \ref{fig:sens}. We also make a brief
comparison with IRAS results at 12$\mu$m.

\subsection{Disk sensitivity at 22-24$\mu$m}\label{ss:det}

\begin{figure*}
  \begin{center}
    \hspace{-0.5cm} \includegraphics[width=0.5\textwidth]{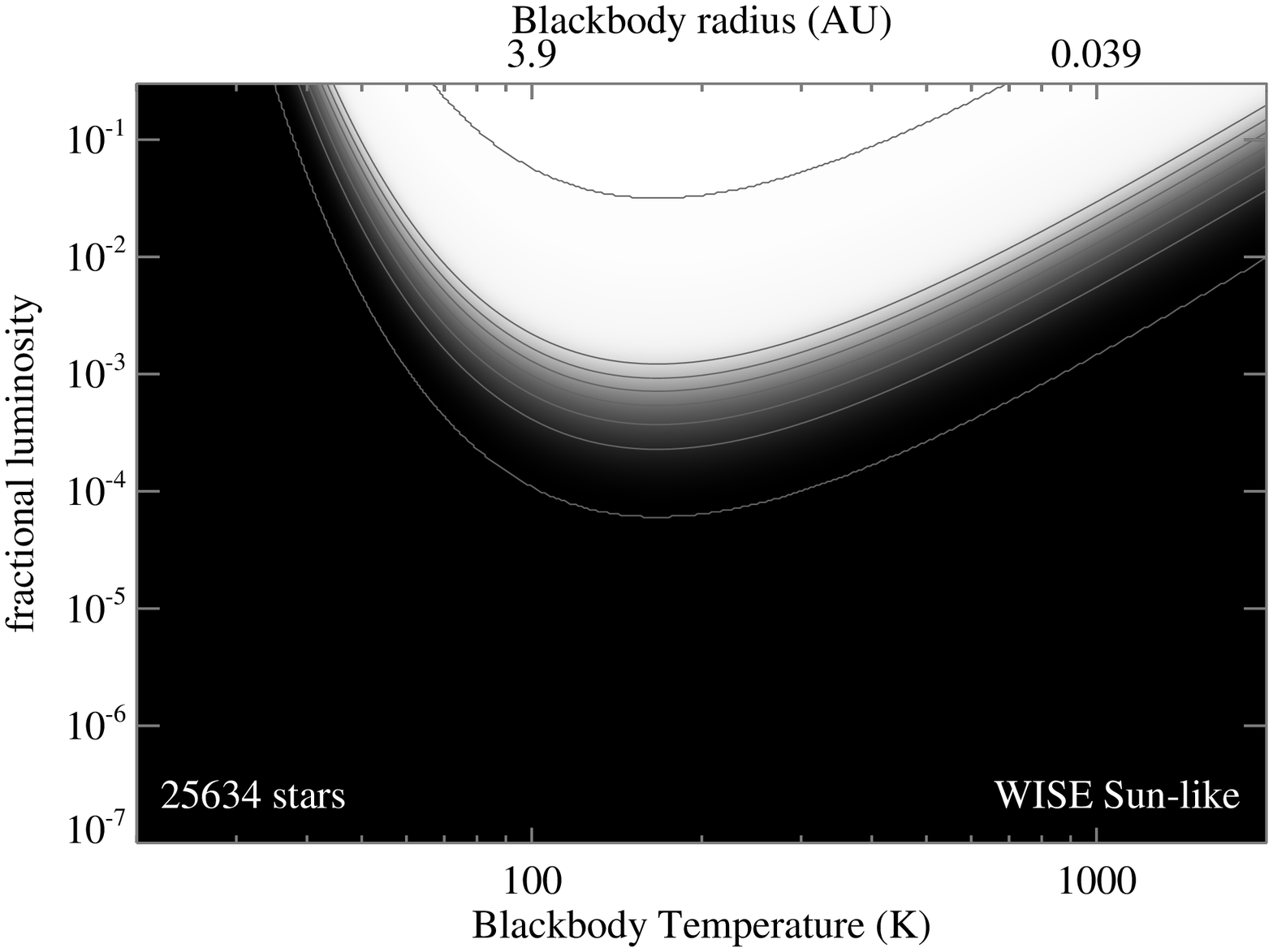}
    \hspace{0.2cm} \includegraphics[width=0.5\textwidth]{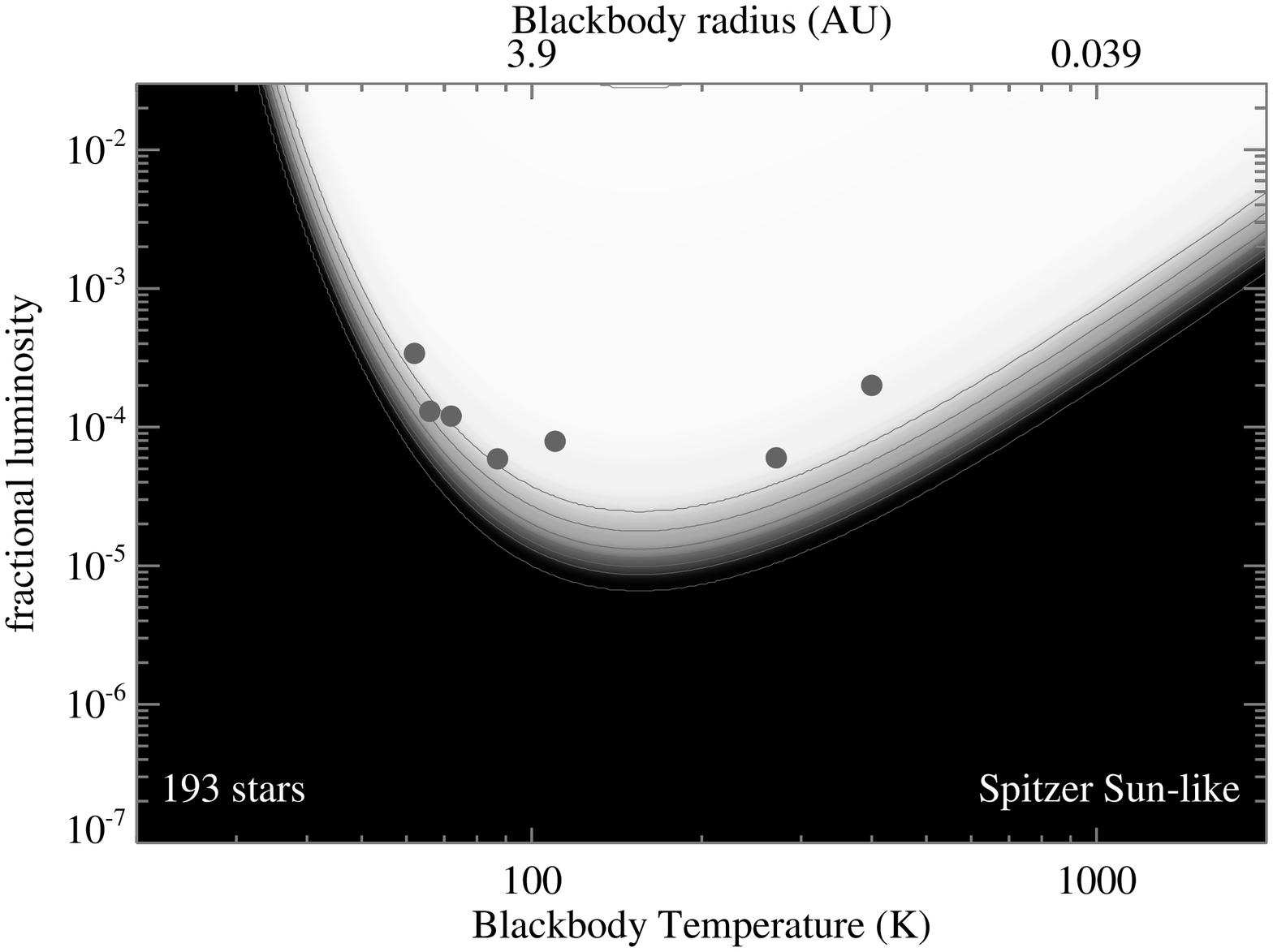}\\
    \vspace{0.2cm}
    \hspace{-0.5cm} \includegraphics[width=0.5\textwidth]{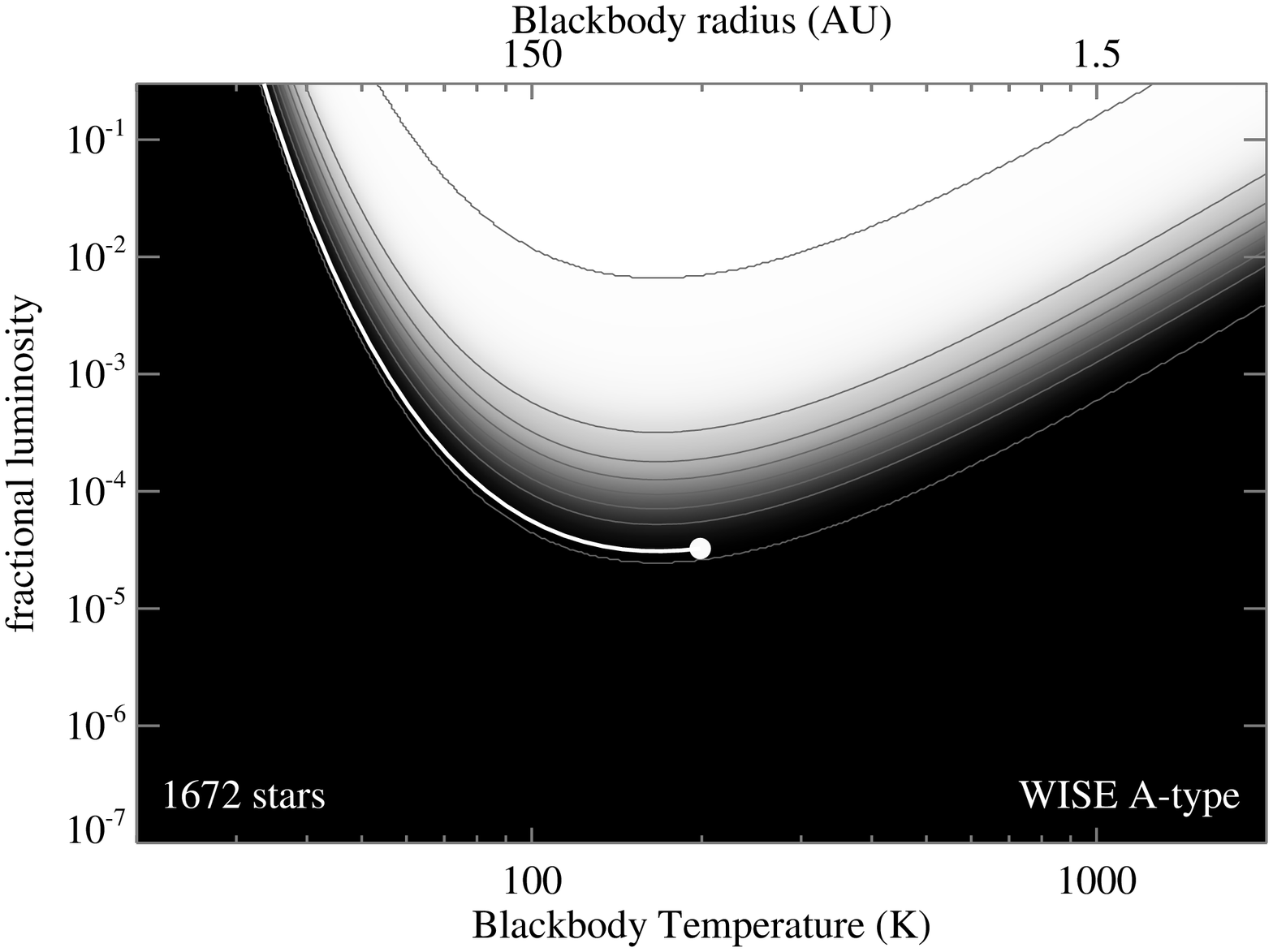}
    \hspace{0.2cm} \includegraphics[width=0.5\textwidth]{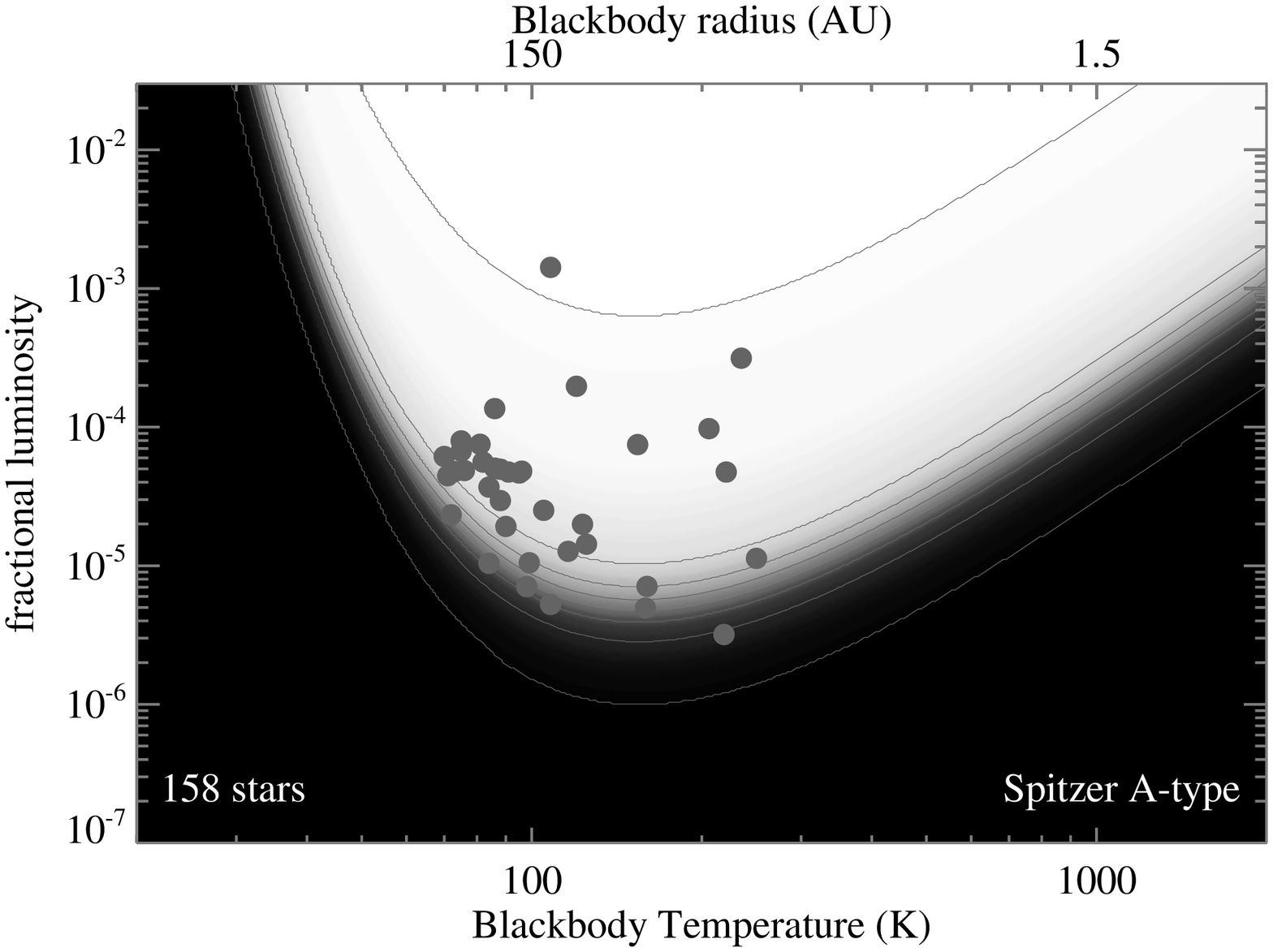}
    \caption{Sun-like (top row) and A-type (bottom row) disk sensitivity comparison
      between WISE 22$\mu$m (left column) and nearby stars with \emph{Spitzer} MIPS at
      24$\mu$m (right column). Disks in regions of the parameter space that are white
      could be detected around all stars, and disks in black regions could not be
      detected around any star. The colour scale is a linear stretch, contours show 8
      linearly spaced levels from 1 to the number of stars observed in each case. The top
      radial scale assumes $L_\star=0.5 L_\odot$ for Sun-like stars and $L_\star=20
      L_\odot$ for A-types. The A-star disk candidate is shown at the temperature fitted
      in Figure \ref{fig:sedw4}, but could lie anywhere along the white line because the
      temperature is only an approximate upper limit.}\label{fig:det}
  \end{center}
\end{figure*}

The left panels of Figure \ref{fig:det} show the sensitivity to disks with WISE at
22$\mu$m, split into Sun-like and A-type stars. The plots are similar to Figure
\ref{fig:sens}, but now represent the cumulative sensitivity for all objects. Disks in
the white region could have been detected around all stars, and disks in the black region
could not have been detected around any star.

For the Sun-like \emph{Kepler} stars observed with WISE (top left panel), the region
covered for the bulk of the stars is similar to that predicted in Figure
\ref{fig:sens}. Only the brightest few stars have sensitivity to fractional luminosities
lower than about 0.1\%. No disks are shown on this plot because the W4 excesses around
Sun-like stars are consistent with arising entirely from background galaxies.

The WISE sensitivity is in contrast to that for nearby Sun-like stars observed with
\emph{Spitzer} at 24$\mu$m \citep[top right panel,][]{2008ApJ...674.1086T}, which could
detect disks with much lower fractional luminosities (i.e. the white region extends to
lower $f$). The WISE sensitivity does not extend into the region where disks were
detected with \emph{Spitzer}, so does not probe the same part of the disk distribution as
the \emph{Spitzer} study.

Compared to nearby A-stars observed with \emph{Spitzer} (lower right panel), the disk and
lowest contours for WISE extend into the region covered by the brightest excesses found
by \citet{2006ApJ...653..675S} (i.e. where $T_{\rm disk} \sim 100$-200K and $f \sim
10^{-3}$-$10^{-4}$). Unlike the Sun-like stars, there is therefore some overlap in the
parts of the disk distributions that are detectable with each survey. For the WISE A-star
disk candidate (dot in lower left panel) we assume the disk properties shown in Figure
\ref{fig:sedw4}. Because this temperature is an approximate upper limit, the disk could
lie anywhere along the white line that curves towards the upper left of the figure
(though cooler disks must have significantly higher fractional luminosities). The WISE
detection is very likely typical based on where it lies relative the known distribution
of A-star excesses.

\subsection{Excess distribution at 22-24$\mu$m}\label{ss:stat24}

\begin{figure*}
  \begin{center}
    \hspace{-0.5cm} \includegraphics[width=0.5\textwidth]{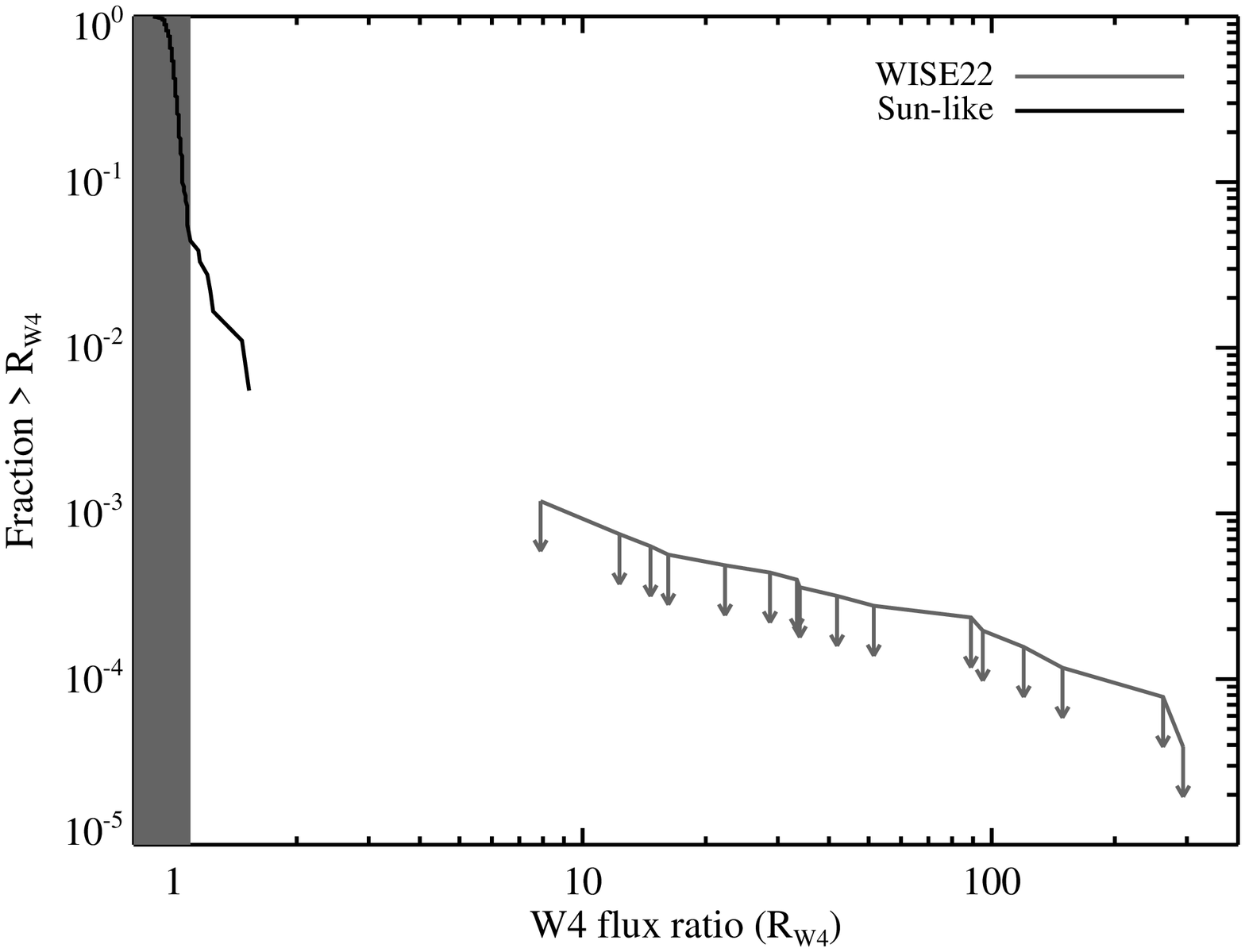}
    \includegraphics[width=0.5\textwidth]{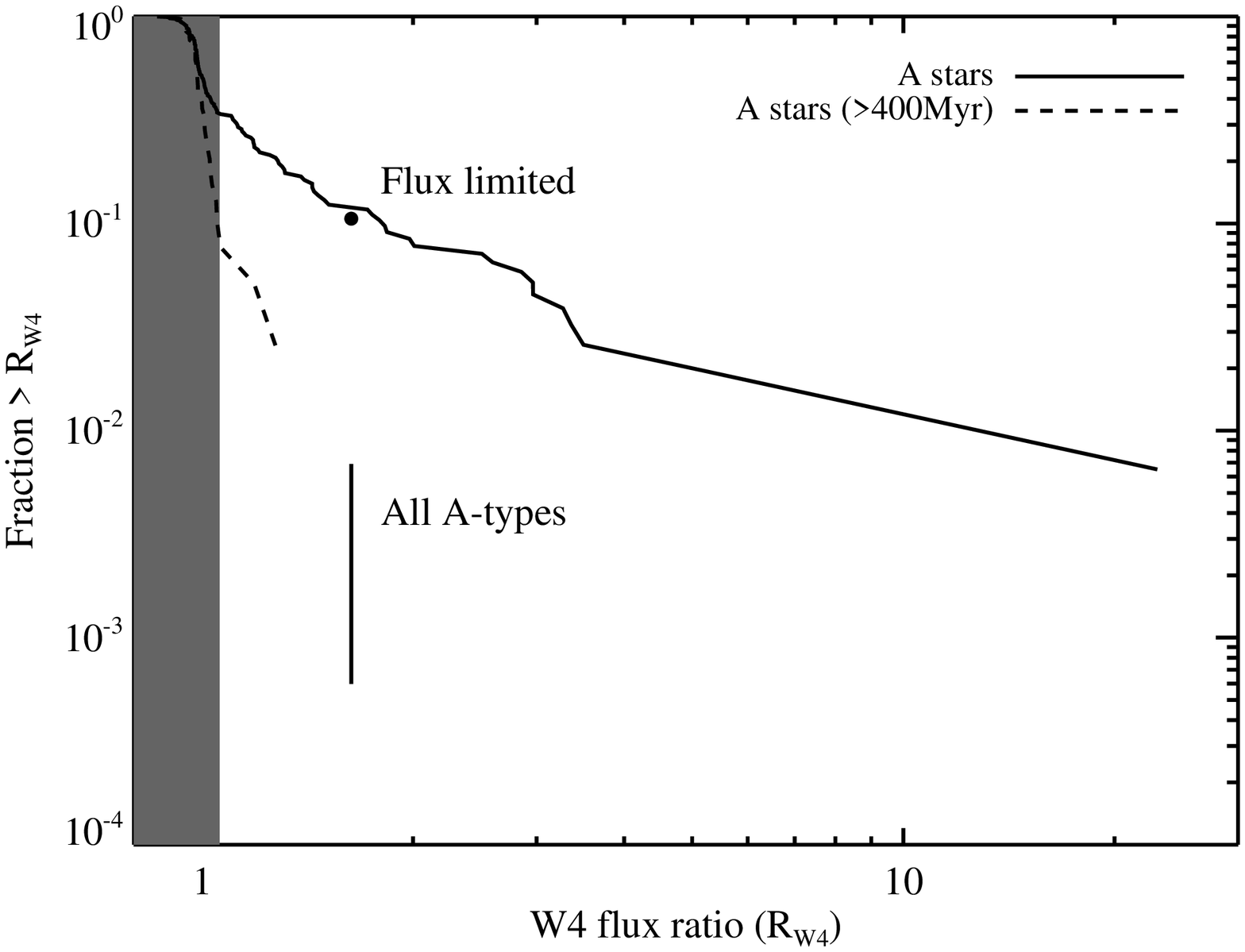}
    \caption{Comparison of 22 and 24$\mu$m flux ratio distributions for WISE and nearby
      stars \citep{2006ApJ...653..675S,2008ApJ...674.1086T}. The left panel shows
      Sun-like stars, for which the WISE distribution is an upper limit. The right panel
      shows A-type stars, and the WISE points are based on the full and flux limited
      samples and the single disk candidate (see text). We also show a distribution for
      nearby A-stars older than 400Myr. The dark regions mark where the \emph{Spitzer}
      observations were calibration limited, below 1.1
      \citep[left,][]{2008ApJ...674.1086T} and below 1.06
      \citep[right,][]{2006ApJ...653..675S}.}\label{fig:cumxs}
  \end{center}
\end{figure*}

Figure \ref{fig:cumxs} shows cumulative 22 and 24$\mu$m flux ratio distributions, again
split into Sun-like and A-type samples. The nearby star distributions are simply the
cumulative distribution of flux ratios, since all observed stars were detected.

For Sun-like stars (left panel), because we concluded that all W4 Sun-like excesses were
consistent with arising from contamination by background galaxies (\S \ref{ss:w34},
Fig. \ref{fig:counts}), the WISE part of the distribution is an upper limit on the
occurrence rate of rare bright disks. It is found by assuming upper limits on flux ratios
are detections (i.e. by assuming that all stars could have disks just below detectable
levels, whereas the true distribution lies somewhere below this level). The lack of
overlap in the distributions due to the rarity of large 22-24$\mu$m excesses, and the
limitations of WISE observations of \emph{Kepler} stars, is clear.

While lower levels of excess (flux ratios of $\sim$1.1-2), have an occurrence rate of
around 2-4\% around Sun-like stars
\citep[e.g. Fig. \ref{fig:cumxs},][]{2006ApJ...638.1070H,2006ApJ...639.1166B}, large
($\gtrsim$2) excesses were previously constrained to less than about 0.5\% based on the
\citet{2008ApJ...674.1086T} sample. We have set new limits 1-2 orders of magnitude lower
and as the left panel of Figure \ref{fig:cumxs} shows, these limits apply to large flux
ratios of 10-300.

For the A-type stars (right panel), we show the excess occurrence rate for the single
A-star with a W4 excess of 1.63. Of the nine stars in the photospheric flux limited
sample without an excess (stars as bright or brighter than the one with an excess), one
has an upper limit higher than 1.63, while the others are lower (i.e. the observations
could have detected a disk like the one found). The occurrence rate at this flux level
therefore lies between 1/9 and 1/10, with these extremes set by assuming that the highest
upper limit is either a detection above 1.63 or unity (a non-detection below 1.63). This
point is shown as ``flux limited'' on Figure \ref{fig:cumxs}, and is consistent with the
sample of \citet{2006ApJ...653..675S}. With only a single detection this occurrence rate
is of course very uncertain.

Considering the full sample of our A-stars yields a lower disk occurrence rate, with at
least 145 stars for which this flux ratio could have been detected. If all upper limits
are assumed to be non-detections below 1.63 then the occurrence is one from 1672
stars. The vertical line in Figure \ref{fig:cumxs} shows the range set by these two
limits, which lies below the point set from the photospheric flux limited sample, and
below the distribution of nearby A-stars.

While these two occurrence rates are very uncertain due to only a single disk detection,
we consider some possible reasons for the discrepancy. One possibility is an age bias, as
most stars in \citet{2006ApJ...653..675S} were chosen based on cluster or moving group
membership and are therefore younger on average then field A-stars.  Cutting the nearby
A-star sample to only contain stars older than 400Myr shows that a difference in sample
ages has a significant effect on the flux ratio distribution \citep[i.e. disks evolve
with time,][]{2005ApJ...620.1010R,2007ApJ...654..580S}. Though the extrapolation of the
$>$400Myr population is very uncertain, this older subsample is more consistent with the
WISE excesses from the full sample. The difference in the distributions could therefore
be understood if A-stars in the \emph{Kepler} field are typically older than about
400Myr. While there should be no such bias for stars of the same spectral type, there is
in fact a difference in the typical spectral types between the \emph{Spitzer} A-star
survey and those observed with WISE. While the \emph{Spitzer} sample comprises late B and
early A-types, our \emph{Kepler} A-stars are mostly at the lower end of the 7000-10,000K
temperature range (i.e. are late A and early F-types). Later spectral types both have
lower disk occurrence rates and are typically older due to longer main-sequence lifetimes
\citep[e.g.][]{2007ApJ...654..580S}, which could account for the lower detection rate. It
is therefore the higher detection rate inferred from the single WISE excess (around a
9700K star) that may be odd, but given the small number (i.e. 1 disk from 10 stars) can
be attributed to chance and that the star with an excess is hotter than most.

\subsection{Excess distribution at 12$\mu$m}\label{ss:stat12}

\begin{figure}
  \begin{center}
    \hspace{-0.5cm} \includegraphics[width=0.5\textwidth]{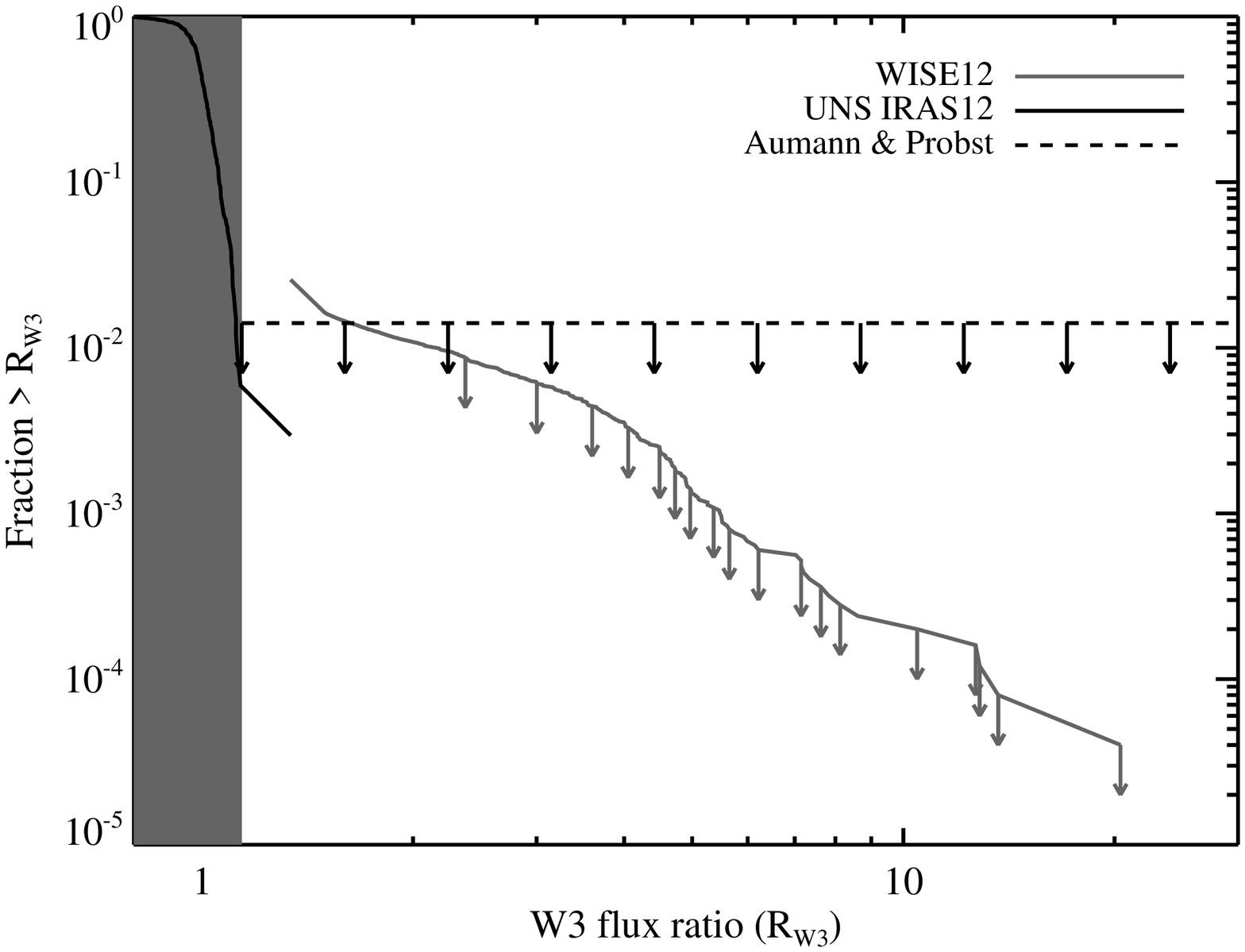}
    \caption{Upper limits on the 12$\mu$m flux ratio distribution from IRAS (dashed line)
      and WISE (solid grey line), compared to the distribution for stars in the UNS
      sample using IRAS. The dark region marks where the IRAS detections were calibration
      limited (below 1.14).}\label{fig:cumxsw3}
  \end{center}
\end{figure}
 
We have also set stringent limits on the distribution of warm disks at 12$\mu$m. At
12$\mu$m, previous knowledge of the excess distribution was derived from the all-sky IRAS
survey \citep[e.g.][]{1991ApJ...368..264A}. While many authors have used the results of
this survey to discover and study warm excesses
\citep[e.g.][]{2005Natur.436..363S,2006ApJS..166..351C,2009ApJ...700L..25M,2010A&A...515A..95S},
few have published the results from an unbiased sample at this wavelength in a manner
that allows the distribution of the 12$\mu$m flux ratios to be determined. Figure
\ref{fig:cumxsw3} shows our upper limit on the 12$\mu$m flux ratio distribution for
Sun-like stars, showing that bright excesses at this wavelength are extremely rare. For
comparison, we show the lack of excess detection among 71 FGK stars detected at 12$\mu$m
\citep{1991ApJ...368..264A}, which was calibration limited and could not detect flux
ratios smaller than 1.14. We also show the distribution of IRAS 12$\mu$m flux ratios for
348 FGK stars in the Unbiased Nearby Star (UNS) sample \citep{2010MNRAS.403.1089P}, based
on photospheric modelling done for the \emph{Herschel} DEBRIS survey
\citep[e.g.][]{2010A&A...518L.135M,2012MNRAS.421.2264K}, which has a similar flux ratio
sensitivity. The only significant excess is for $\eta$ Corvi (HD 69830 is the second
largest excess, but with a flux ratio of 1.13 is only about 2$\sigma$
significant).\footnote{Several other stars in this sample show significant excesses, but
  these can be shown to be spurious based on more recent \emph{Spitzer} MIPS and
  \emph{Herschel} PACS observations that resolve the star and a nearby background
  source.} These constraints and detections are all consistent, and set limits on the
rarity of bright 12$\mu$m excesses to less than one in every thousand to ten thousand
stars for flux ratios greater than about 5.

\section{Discussion}\label{s:disc}

One of several goals for this study was to test for a correlation between the existence
of debris disks and planets discovered by \emph{Kepler}. However, the distribution of the
rare bright excesses that WISE is sensitive to around \emph{Kepler} stars was not known
at the outset, so whether this goal was possible was not known either. We noted that even
if bright disks were too rare among the bulk population, that a possible correlation
between disks and low-mass planets may allow robust disks detections among this subset.

Only one \emph{Kepler} planet candidate host (of 348 KOIs that were not excluded by the
100$\mu$m background cut) was found to have an excess, so this possibility appears
unlikely. In addition, Figure \ref{fig:counts} shows that this detection rate is close to
that expected from galaxy confusion. Thus, for the bright warm excesses that WISE is
sensitive to, there is no evidence that planet host candidates have a disk occurrence
rate that is different from the bulk population.

Similarly, excesses around the remaining \emph{Kepler} stars are also consistent with
arising from chance alignments with background galaxies, with the exception of a single
A-type star. However, the possibility that a small number of the excesses are true debris
disks means that the chance of detecting transiting dust concentrations is at least as
good as for \emph{Kepler} stars without excesses, and may be higher (the 271 Kepler stars
with W3 or W4 excesses are listed in Table \ref{tab:xs}). Discovery of many such dust
transits that preferentially occur around stars with excesses would argue that at least
some excesses are debris disks, though this method of verification seems unlikely.

We have therefore set new limits on the distribution of warm excesses. The range of flux
ratios for which we have set limits for Sun-like stars is 2-20 at 12$\mu$m
(Fig. \ref{fig:cumxsw3}) and 10-300 at 22$\mu$m (left panel of Fig. \ref{fig:cumxs}). For
such large 12-22$\mu$m excesses to arise from steady-state processes the planetesimal
belts would have to be either around very young stars or relatively distant from their
central star \citep{2007ApJ...658..569W}, which in turn requires fractional luminosities
$\gtrsim$1\% (see Fig. \ref{fig:sens}). Detecting large warm excesses around
main-sequence stars is very unlikely because collisional evolution depletes belts near
the central star to undetectable levels rapidly, so the conclusion is that such mid-IR
excesses are most likely transient. Two main processes seem to be plausible causes of
such excesses. The first, delivery of material from an outer reservoir
\citep{2005ApJ...626.1061B,2007ApJ...658..569W}, is appealing because short-lived warm
dust can be replenished using material from a long-lived outer belt. Alternatively,
because we are here interested in large excesses, the debris from a giant impact between
large bodies is a possibility \citep[i.e. perhaps similar to the Earth-Moon forming
event,][]{2012arXiv1206.4190J}.

Several possibilities exist for the delivery of objects from an outer belt to terrestrial
regions. A system of sufficiently many planets on stable orbits can pass objects inwards
from an outer belt \citep{2012MNRAS.420.2990B}, or a planetary system instability can
severely disturb a planetesimal population, some of which end up in the terrestrial zone
\citep{2005Natur.435..466G}. Such possibilities have been suggested as mechanisms to
generate the warm dust component observed around $\eta$ Corvi
\citep{2009MNRAS.399..385B,2011arXiv1110.4172L}.

Because at least 15\% of Sun-like stars have cool outer planetesimal belts
\citep[e.g.][]{2008ApJ...674.1086T}, our limits of 0.01-0.1\% for warm belts (for the
flux ratios noted above for 12 and 22$\mu$m) mean that fewer than 1 in 150-1500 can be
generating large levels of warm dust from cool outer belts at any given time. This
fraction could in fact be larger because the 15\% only represents disks down to a
particular detection limit, and cool disks too faint to detect could still have enough
material to produce large warm dust levels
\citep{2007ApJ...658..569W}. \citet{2009MNRAS.399..385B} placed similar limits on the
number of systems that could be caught in the act of an instability that delivers large
amounts of debris to the terrestrial region, estimating that less than about 0.2\%
(i.e. 1/500) of Sun-like stars might be observed undergoing an instability at 24$\mu$m.

Whether such instabilities do produce very large excesses is another question. In
studying the dust emission generated in their model of the Solar System's proposed
planetary instability, \citet{2009MNRAS.399..385B} find that while the relative changes
can be very large, the flux ratios are near unity at 12$\mu$m and of order 10 at
24$\mu$m. However, these ratios may be underestimated because they do not incude emission
that could arise from the sublimation of comets within 1AU. It is therefore hard to say
whether these results are representative, since they will also depend on the specific
system architecture. The $\eta$ Corvi system has been suggested as a possible candidate
currently undergoing such an instability, and shows a 12$\mu$m flux ratio of 1.3. If
typical, these results suggest that instabilities may not produce the larger excesses
considered here.

In contrast, the giant impact scenario can produce extremely large excesses
\citep{2012arXiv1206.4190J}. The relatively nearby star BD+20 307 (at 96pc), which has a
10$\mu$m flux ratio of about 100, is a good candidate for such an event
\citep{2005Natur.436..363S,2011ApJ...726...72W}. While such events would generally be
expected to be associated with young systems, where the final $\sim$10-100Myr chaotic
period of giant impacts and terrestrial planet formation is winding down
\citep[e.g.][]{1998Icar..136..304C,2001Icar..152..205C}, BD+20 307 is a $\gtrsim$Gyr old
main-sequence binary \citep{2008ApJ...688.1345Z}. The excess may therefore be indicative
of a recent instability that has greatly increased the chance of collisions within the
terrestrial zone, and is unrelated to planet formation
\citep{2008ApJ...688.1345Z}. Clearly, age estimates for the host stars are important for
understanding the origin of dust in such systems.

While the WISE mission might appear to permit near-unlimited sample sizes to help detect
the aftermath of the rarest collision events, we have shown that their detection among
\emph{Kepler} stars is fundamentally limited. This limit arises because the occurrence
rate of excesses that can be detected is too low, so the disks are overwhelmed by galaxy
contamination. Because \emph{Kepler} stars represent a sample that will remain unique for
the forseeable future, it is desirable to find ways to overcome this issue. Based on the
findings of \S \ref{ss:w34}, one option is to create sub-samples that maximise the chance
of disk detection, because higher disk occurrence rates are more robust to galaxy
contamination. Younger stars tend to have larger excesses that are also more frequent
\citep[e.g.][]{2005ApJ...620.1010R,2007ApJ...654..580S,2009ApJS..181..197C}, so a
sub-sample of young stars will be more robust to confusion. The long-term monitoring of
\emph{Kepler} stars may provide some help if accurate stellar ages can be derived, for
example if rotation periods can be derived to yield age estimates via gyrochronology
\citep{1972ApJ...171..565S,2007ApJ...669.1167B,2008ApJ...687.1264M}. Another way to split
the sample is by spectral type, because earlier-type stars are both brighter and have
higher disk occurrence rates (for fixed sensitivity). This approach is less appealing for
studying the links between disks and planets however, because the bulk of stars observed
by \emph{Kepler} are Sun-like.

If we allow for the possibility of observing \emph{Kepler} stars with WISE excesses with
other instruments, there is a potential gain with better resolution. A galaxy that is
unresolved with WISE might be resolved with \emph{Spitzer}'s IRAC instrument, or using
ground-based mid-IR observations on 8m-class telescopes for example. Assuming that it
could be detected, the high ($\sim$0\farcs5) resolution of such ground-based observations
would have over a 99\% chance of detecting a galaxy that was not resolved with the WISE
beam at 12$\mu$m. Therefore, detection of fewer galaxies than expected in a sample of
targets (e.g. significantly fewer than 99 out of 100) would be evidence that the excesses
do not randomly lie within the WISE beam and that some are therefore due to excesses
centered on the star (i.e. are debris disks).

Ultimately, we found that searching for debris disks around stars in the \emph{Kepler}
field with WISE is limited by the high background level and galaxy contamination. While
high background regions can be avoided, background galaxies will always be an issue for
such distant stars. Though it means being unable to study the planet-disk connection with
such a large planet-host sample, nearby stars should be the focus of studies that aim to
better define the distribution of warm excesses. Characterising this distribution is very
important, particularly for estimating the possible impact of terrestrial-zone dust on
the search for extrasolar Earth analogues
\citep[e.g.][]{2006ApJ...652.1674B,2012arXiv1204.0025R}. For example, extending the
distribution to the faintest possible level available with photometry (calibration
limited to a 3$\sigma$ level of $\sim$5\%) yields a starting point to make predictions
for instruments that aim to detect faint ``exozodi'' with smaller levels of
excess. Because bright warm debris disks must decay to (and be observable at) fainter
levels, the distribution will also provide constraints on models that aim to explain the
frequency and origin of warm dust.

\section{Summary}\label{s:sum}

We have described our search of about 180,000 stars observed by \emph{Kepler} for debris
disks using the WISE catalogue. With the completion of the AKARI and WISE missions, such
large studies will likely become common. We have identified and addressed some of the
issues that will be encountered by future efforts, which mainly relate to keeping
spurious excesses to a minimum by using information provided in photometric catalogues.

We used an SED fitting method to identify about 8000 infra-red excesses, most of which
are in the 12$\mu$m W3 band around Sun-like stars. The bulk of these excesses arise due
to the high mid-IR background level in the \emph{Kepler} field and the way source
extraction is done in generating the WISE catalogue. From comparing the number counts for
excesses in low background regions with cosmological surveys and WISE photometry from the
\emph{Kepler} field, we concluded that a 22$\mu$m excess around a single A-type star is
the most robust to confusion, with about a 1/100 chance of arising due to a background
galaxy. We found no evidence that the disk occurrence rate is any different for planet
and non-planet host stars.

In looking for these disks we have set new limits on the occurrence rate of warm bright
disks. This new characterisation shows why discovery of rare warm debris disks around
Sun-like \emph{Kepler} stars in low background regions is generally limited by galaxy
confusion. Though the planetary aspect would be lost, nearer stars should be the focus of
future studies that aim to characterise the occurrence of warm excesses.

\section*{Acknowledgements}

We thank the reviewer for a thorough reading of this article and valuable comments. This
work was supported by the European Union through ERC grant number 279973. This research
has made use of the following: The NASA/ IPAC Infrared Science Archive, which is operated
by the Jet Propulsion Laboratory, California Institute of Technology, under contract with
the National Aeronautics and Space Administration. Data products from the Wide-field
Infrared Survey Explorer, which is a joint project of the University of California, Los
Angeles, and the Jet Propulsion Laboratory/California Institute of Technology, funded by
the National Aeronautics and Space Administration. Data products from the Two Micron All
Sky Survey, which is a joint project of the University of Massachusetts and the Infrared
Processing and Analysis Center/California Institute of Technology, funded by the National
Aeronautics and Space Administration and the National Science Foundation. The
Multimission Archive at the Space Telescope Science Institute (MAST). STScI is operated
by the Association of Universities for Research in Astronomy, Inc., under NASA contract
NAS5-26555. Support for MAST for non-HST data is provided by the NASA Office of Space
Science via grant NNX09AF08G and by other grants and contracts. Data collected by the
Kepler mission. Funding for the Kepler mission is provided by the NASA Science Mission
directorate.



\onecolumn

\clearpage

\appendix

\section{Stars with excesses}

\begin{longtable}{lrrrrrrrl}
  \caption{The 271 \emph{Kepler} stars with WISE 3-4 excesses. Columns are: KIC
    identifier, predicted \emph{Kepler} ($K_p$) magnitude from the KIC, Quarters the star
    was observed in, fitted effective temperature, W3-4 flux ratio and excess
    significance (where $X_{\rm W3,4} \ge 4$). The note column notes the single KOI,
    and potential planet hosts from
    \citet{2012arXiv1201.1048T} (``T12'').}\label{tab:xs1}\\
  KIC & $K_p$ & Quarters & $T_{\rm eff}$ & $R_{\rm W3}$ & $X_{\rm W3}$ & $R_{\rm W4}$ & $X_{\rm W4}$ & Notes \\
  \hline
    \endfirsthead
    KIC & $K_p$ & Quarters & $T_{\rm eff}$ & $R_{\rm W3}$ & $X_{\rm W3}$ & $R_{\rm W4}$ & $X_{\rm W4}$ & Notes \\
    \hline
    \endhead
    5866211&15.19&456&6585&4.2&5.0&&&\\
5866341&15.06&123456&6296&4.9&6.7&&&\\
5866415&15.33&123&6029&4.7&4.4&&&\\
6198278&14.86&123456&5436&2.8&4.1&&&\\
6346886&14.96&12346&5869&3.9&4.6&&&\\
6431431&14.87&123456&8147&5.9&7.8&&&\\
6503763&15.78&12346&5275&4.3&4.1&&&\\
6515382&13.29&123456&6265&1.7&4.2&&&\\
6516101&13.88&123456&6062&2.3&6.1&&&\\
6599949&15.42&123456&5773&4.0&4.1&&&\\
6676683&14.58&123456&6356&4.1&6.2&&&\\
6685526&15.00&123456&5103&2.7&4.1&&&KOI 861,T12\\
6773853&14.89&123456&6041&2.9&4.0&&&\\
6935614&15.73&123456&5832&7.1&7.1&&&\\
7022341&15.52&456&6111&5.1&4.4&&&\\
7104629&15.41&123456&5824&4.9&4.6&&&\\
7104793&15.51&123456&5013&3.6&4.8&&&\\
7184587&15.61&123456&3963&&&33.2&4.1&\\
7187014&15.46&123456&5940&4.9&5.1&&&\\
7187096&15.07&123456&4153&2.0&4.9&&&\\
7189185&15.18&123456&5359&4.9&5.9&&&\\
7268366&14.52&123456&6034&4.6&7.7&&&\\
7345479&7.93&123456&9686&&&1.6&4.5&\\
7349062&14.90&123456&6606&3.8&4.7&&&\\
7349090&14.76&123456&6173&4.1&6.2&&&\\
7350204&14.85&1256&6385&3.7&4.6&&&\\
7354462&15.29&123456&6388&4.5&4.5&&&\\
7516798&15.10&123456&5582&3.1&4.0&&&\\
7581686&12.61&123456&6342&1.9&7.4&14.7&5.8&\\
7593434&14.92&123456&5559&3.3&5.1&&&\\
7595932&13.44&123456&5039&2.5&9.8&&&\\
7597096&15.78&256&6273&6.2&4.7&&&\\
7659091&13.57&123456&5844&1.8&4.5&&&\\
7667940&14.48&123456&5716&3.3&6.2&&&\\
7673565&15.62&23&6363&5.5&4.7&&&\\
7730130&16.17&2&4860&4.9&4.6&&&T12\\
7731810&15.55&2&7482&3.8&4.2&&&\\
7744202&15.52&456&5246&4.0&4.6&&&\\
7744209&15.70&123456&5143&5.0&5.1&&&\\
7746956&15.27&123456&6976&4.7&5.0&&&\\
7808214&15.49&123456&4611&4.3&9.6&&&\\
7811074&15.43&123456&5742&4.5&4.5&&&\\
7877878&15.86&456&6026&7.2&4.7&&&\\
7877962&14.46&123456&5937&2.4&4.1&&&\\
7879639&15.90&123456&4896&5.5&5.7&&&\\
8005470&14.18&123456&5608&2.0&4.3&&&\\
8013236&15.72&23456&5241&4.6&4.6&&&\\
8016698&13.45&123456&7910&2.0&4.9&&&\\
8075618&15.67&123456&5666&4.5&4.3&&&T12\\
8077083&15.92&123456&5456&12.7&10.2&293.7&9.2&\\
8085263&15.76&23456&6369&12.9&8.5&&&\\
8145154&14.09&123456&6144&2.2&4.8&&&\\
8145181&14.98&123456&5240&2.6&4.5&&&\\
8153997&15.14&123456&5841&3.5&4.8&&&\\
8212592&16.66&2&3840&3.6&5.5&&&\\
8213938&13.14&123456&5650&1.8&5.8&&&\\
8284699&15.42&123456&5303&3.8&4.1&&&\\
8284814&14.92&123456&6368&3.2&4.3&&&\\
8345414&15.29&123456&4184&2.0&4.0&&&\\
8349926&14.78&123456&5819&2.6&4.2&&&\\
8350421&15.69&123456&4674&2.7&4.3&&&\\
8351168&13.93&123456&5230&2.3&7.5&&&\\
8410210&15.36&123456&5957&4.7&4.1&&&\\
8410749&15.01&456&6608&4.9&7.3&&&\\
8417035&15.84&456&4680&3.4&4.0&&&\\
8607558&15.29&123456&5937&3.7&4.4&&&\\
8611027&15.56&23456&4869&3.5&5.0&&&\\
8612202&14.56&123456&8504&2.6&4.4&&&\\
8612850&15.29&123456&6206&5.3&6.2&&&\\
8672241&14.03&123456&5852&2.2&5.0&&&\\
8736331&15.86&123456&6269&7.4&5.9&&&\\
8736639&14.73&123456&5092&2.2&4.2&&&\\
8741807&15.15&123456&6855&4.2&4.6&&&\\
8800998&13.72&123456&9000&6.7&10.3&&&\\
8803050&13.92&123456&6376&&&34.0&4.4&\\
8807242&14.31&123456&5850&2.7&4.5&&&\\
8870902&13.55&123456&6342&2.4&5.5&&&\\
9071384&15.64&123456&4734&3.3&4.9&&&\\
9074768&14.72&123456&5915&8.1&12.9&89.0&6.1&\\
9074812&14.53&123456&5961&2.3&4.2&&&\\
9076617&14.49&123456&5921&2.5&4.3&&&\\
9137443&13.76&123456&5879&1.9&4.6&&&\\
9138286&14.15&123456&5603&2.1&4.3&&&\\
9139782&15.76&456&5949&5.0&5.0&&&\\
9142411&15.19&123456&5953&4.2&4.0&&&\\
9206761&15.82&23456&6021&7.2&5.0&&&\\
9264468&14.36&123456&5440&2.1&4.3&&&\\
9267353&14.78&123456&6200&3.8&6.4&&&\\
9269492&14.64&123456&4856&2.4&5.8&&&\\
9328535&13.87&123456&5900&1.8&4.6&&&\\
9329967&13.35&123456&6356&2.8&9.3&&&\\
9452213&13.40&123456&8410&1.7&4.8&&&\\
9511303&14.95&123456&5975&2.9&4.3&&&\\
9511944&14.12&123456&6364&2.3&4.3&&&\\
9512868&13.86&23456&5289&2.3&6.5&&&\\
9575361&15.86&123456&4139&4.1&7.8&&&\\
9691491&15.75&123456&4117&2.3&4.0&&&\\
9703058&16.77&5&3995&4.4&5.2&&&\\
9762054&15.84&456&6044&5.6&4.2&&&\\
9813767&15.99&123456&5039&4.8&4.9&&&\\
9823991&15.74&123456&5520&5.4&5.0&&&\\
9824039&14.99&123456&5487&3.5&4.1&&&\\
9873729&13.55&123456&3903&1.3&4.4&&&\\
9873862&15.28&123&5040&2.8&4.5&&&\\
9875170&15.28&123456&5469&2.9&4.3&&&\\
9875827&15.36&123456&5114&5.7&8.7&&&\\
9883553&12.93&123456&6586&1.6&4.5&&&\\
9883654&14.96&123456&6351&3.2&4.3&&&\\
9883689&14.74&123456&5409&2.7&5.4&33.4&4.0&\\
9883939&14.82&123456&5852&3.0&4.4&&&\\
9933368&15.03&123456&6037&5.5&7.0&&&\\
9933625&12.82&123456&5550&1.5&5.0&&&\\
9936573&14.62&123456&5527&2.4&4.5&&&\\
10002543&13.24&123456&7313&1.7&4.5&&&\\
10002794&15.96&23456&5259&7.0&5.1&&&\\
10056410&15.24&123456&5778&3.5&4.3&&&\\
10062742&15.77&23&6370&5.6&5.5&&&\\
10063763&13.31&123456&7532&2.0&6.1&&&\\
10065701&14.51&123456&6145&2.6&4.7&&&\\
10119646&15.74&123456&5418&4.7&4.9&&&\\
10120908&15.83&123456&4915&4.3&5.3&&&\\
10128226&15.16&123456&6157&4.5&5.4&&&\\
10128466&13.18&1&6249&1.7&4.2&&&\\
10128553&15.18&123456&5428&4.1&5.2&&&\\
10128580&15.44&123456&5208&3.7&4.3&&&\\
10128587&15.17&123456&4813&2.5&4.7&&&\\
10131814&14.15&123456&5415&2.1&4.9&&&\\
10192175&15.12&123456&5624&3.9&5.3&&&\\
10195974&14.96&123456&5539&3.0&4.4&&&\\
10199239&13.66&123456&5605&1.9&4.4&&&\\
10199401&15.49&123&5556&6.1&7.7&&&\\
10252275&13.75&123456&5132&1.5&4.1&&&\\
10252286&14.94&123456&6057&4.0&6.1&&&\\
10252364&13.55&123456&5984&1.7&4.2&&&\\
10253878&15.86&456&6377&7.8&6.7&&&\\
10255817&15.37&123456&5071&4.7&6.7&&&\\
10256442&15.48&123456&5469&3.7&4.4&&&\\
10256507&14.91&123456&5119&2.5&4.7&&&\\
10264259&14.92&123456&4967&2.6&6.0&&&\\
10265238&15.08&123456&5767&2.8&4.7&&&\\
10265241&15.11&123456&5631&3.1&4.9&&&\\
10265602&15.55&23456&5855&4.8&5.0&&&\\
10318128&14.80&123456&6628&4.0&5.8&&&\\
10321367&15.22&123456&6176&4.7&5.4&&&\\
10321406&14.95&123456&5853&3.3&5.2&&&\\
10321407&15.30&456&6306&3.9&4.2&&&\\
10321422&15.06&123456&4832&2.5&4.8&&&\\
10322187&15.69&23&5578&4.7&4.1&&&\\
10322220&15.30&456&5028&2.8&4.7&&&\\
10328472&14.72&123456&5934&2.7&4.1&&&\\
10330579&14.53&123456&6281&2.4&4.4&&&\\
10382415&15.63&56&4073&3.6&8.9&&&\\
10383222&15.51&23456&4959&2.9&4.1&&&\\
10386716&16.86&2&4844&10.5&6.1&&&\\
10386900&14.88&123456&5583&2.7&4.3&&&\\
10387564&15.35&123456&5909&4.0&4.6&&&\\
10395762&15.50&123456&6072&4.1&4.0&&&\\
10395814&15.89&456&4806&&&95.1&4.4&\\
10447798&14.74&123456&5241&2.4&5.0&&&\\
10451070&15.00&123456&5038&2.4&4.0&&&\\
10451135&14.24&123456&5354&2.3&5.2&&&\\
10451251&15.66&456&4705&3.1&5.2&&&\\
10451497&15.20&123456&5460&3.2&4.7&&&\\
10451632&15.33&123456&4771&2.8&5.3&&&\\
10461970&15.77&123456&5645&5.5&4.8&&&\\
10513479&14.97&123456&8410&6.4&9.8&&&\\
10513812&14.41&123456&6256&2.5&4.8&&&\\
10515235&15.53&23456&5924&5.0&5.1&&&\\
10515276&15.26&123456&3917&2.2&7.4&&&\\
10516147&14.84&123456&6136&3.6&6.5&&&\\
10516255&15.34&123456&5948&4.1&5.2&&&\\
10517028&13.48&123456&6630&2.0&4.8&&&\\
10517486&15.35&1256&5050&2.9&4.4&&&\\
10579318&15.42&123456&5304&3.5&4.3&&&\\
10580355&14.78&123456&5473&3.2&5.9&&&\\
10580447&15.84&123456&3987&2.7&5.8&&&\\
10580525&13.86&123456&6388&2.0&4.6&&&\\
10580779&15.64&123456&5470&4.5&5.7&&&\\
10580786&15.38&123&5897&5.0&5.9&&&\\
10581163&13.60&123456&5973&1.7&4.4&&&\\
10581308&13.77&123456&5460&1.8&5.1&&&\\
10581836&15.39&123456&4841&3.8&6.8&&&\\
10582786&14.68&123456&5765&2.3&4.3&&&\\
10583400&14.86&123456&5217&2.4&4.6&&&\\
10583563&14.11&123456&5482&1.8&4.3&&&\\
10591195&14.67&123456&5941&3.6&6.5&&&\\
10644697&14.17&123456&6396&2.3&5.3&&&\\
10645900&15.83&145&5624&4.2&4.0&&&\\
10645926&14.33&123456&6393&2.2&4.1&&&\\
10646068&17.51&6&3813&5.5&4.3&&&\\
10646091&15.04&123456&4117&1.8&4.8&&&\\
10646106&15.33&123456&5585&3.4&4.6&&&\\
10646263&15.56&123456&6091&5.9&5.5&&&\\
10646283&15.40&123456&5465&3.9&4.7&&&\\
10646426&14.69&123456&4768&1.8&4.3&&&\\
10646589&14.62&123456&5149&2.5&5.4&&&\\
10649444&15.37&456&3900&1.6&4.1&&&\\
10649541&15.63&123456&5704&4.2&4.9&&&\\
10649562&14.14&123456&5436&2.1&5.5&&&\\
10656438&15.99&123456&4574&3.0&4.7&&&\\
10710753&15.50&123456&4044&1.9&4.3&&&\\
10711021&15.96&123456&4852&4.1&5.0&&&\\
10711045&15.88&123456&4857&3.4&4.3&&&\\
10711052&15.53&123456&4924&4.5&7.1&&&\\
10711088&14.03&6&6153&2.4&5.0&&&\\
10711259&15.16&123456&6207&3.4&4.0&&&\\
10711510&15.02&123456&6210&3.2&4.1&&&\\
10714422&15.33&123456&4759&2.4&4.6&&&\\
10714459&15.58&23&6000&6.0&6.4&&&\\
10714581&13.95&145&4388&1.5&5.5&&&\\
10716598&14.37&123456&5350&2.0&4.0&&&\\
10721855&13.86&123456&5212&&&16.2&4.0&\\
10722535&15.84&123456&5348&5.5&5.1&&&\\
10724544&15.01&123456&5546&5.1&8.0&&&\\
10777410&15.09&123456&3900&1.4&4.4&&&\\
10777448&15.71&123456&4067&2.4&5.4&&&\\
10777591&13.33&123456&5517&2.2&8.1&&&\\
10777728&15.15&123456&6080&3.6&4.5&&&\\
10778016&15.24&123456&5919&3.8&4.5&&&\\
10845333&15.64&123456&5221&3.7&4.8&&&\\
10907059&14.50&123456&6276&2.5&4.1&&&\\
10907132&15.68&23456&6167&4.6&4.3&&&\\
10908054&13.44&123456&6353&2.0&5.4&&&\\
10958951&12.67&123456&6459&2.4&9.9&12.3&4.4&\\
11017907&11.80&123456&5873&&&7.9&6.7&\\
11074521&15.57&123456&4921&3.7&5.7&&&\\
11075222&14.33&123456&5593&2.4&5.8&&&\\
11086203&14.30&123456&6543&2.9&6.1&&&\\
11086359&15.83&456&6052&4.7&4.1&&&\\
11125136&15.33&123456&5767&4.5&5.5&&&\\
11180691&15.94&456&5065&4.0&4.1&&&\\
11181653&15.25&123456&5936&3.8&4.6&&&\\
11190125&13.75&123456&6737&1.9&4.4&&&\\
11288574&13.46&123456&5408&1.7&5.6&&&\\
11296807&15.06&123456&6036&4.3&6.3&&&\\
11341446&13.42&123456&5553&1.6&4.5&&&\\
11393569&15.99&123456&4563&3.3&4.8&&&\\
11401060&15.36&123456&7981&12.3&13.0&&&\\
11401954&13.21&123456&5870&1.6&4.3&&&\\
11404100&15.00&123456&5045&2.3&4.5&41.9&5.2&\\
11442840&15.07&123456&5322&3.5&5.6&&&\\
11444855&15.10&145&5642&4.2&6.3&&&\\
11456355&13.57&123456&7298&1.8&4.4&&&\\
11457002&15.94&2&5889&4.7&5.6&&&T12\\
11457020&14.90&123456&6120&2.7&4.4&&&\\
11457038&14.32&123456&5744&2.2&5.1&&&\\
11493473&13.63&123456&6125&3.2&10.3&22.3&4.8&\\
11493497&15.41&123456&3897&2.3&8.5&&&\\
11507003&12.67&123456&5467&1.3&4.2&&&\\
11507053&15.84&456&5565&5.3&5.7&&&\\
11507127&14.27&123456&5925&3.0&7.3&28.7&4.1&\\
11507139&15.94&2456&5979&8.6&7.2&&&\\
11546374&15.74&123456&3854&1.9&4.9&&&\\
11546397&15.99&123456&4758&3.3&4.3&&&\\
11551210&12.83&123456&5863&3.3&9.0&&&\\
11558249&15.17&123456&6210&3.6&4.5&&&\\
11598638&13.73&123456&6165&3.5&9.1&&&\\
11649347&15.93&123456&5043&7.6&8.3&148.9&5.9&\\
11649744&13.88&123456&5014&2.1&7.4&&&\\
11700640&15.24&123456&6151&4.2&4.6&&&\\
11753371&15.42&123456&5527&4.4&6.5&&&\\
11803544&15.14&123456&5038&2.7&4.8&&&\\
11855348&15.21&456&6286&13.7&15.2&119.7&5.9&\\
11855417&13.80&123456&5841&1.8&4.2&&&\\
11903173&14.99&12356&4938&3.0&6.3&&&\\
12053791&15.46&123456&5721&3.9&4.4&&&\\
12056198&15.68&123456&5577&20.4&15.6&262.3&9.1&\\
12058865&14.42&123456&6125&3.1&6.1&51.5&4.9&\\
12058904&14.88&145&5095&3.4&7.1&&&\\
12350553&14.53&123456&5935&3.5&7.1&&&\\
12506956&15.09&123456&5030&2.7&4.9&&&\\
\end{longtable}

\end{document}